\documentclass[
reprint,
superscriptaddress,
amsmath,amssymb,
aps,
prb
]{revtex4-2}

\usepackage{graphicx} % Figures
\usepackage{dcolumn}  % Align table columns on decimal point
\usepackage{bm}       % Bold math symbols
\usepackage[colorlinks=true, linkcolor=blue, citecolor=blue, urlcolor=blue]{hyperref} % Hyperlinks
\newcommand{\orcid}[1]{\href{https://orcid.org/#1}{\includegraphics[height=1.6ex]{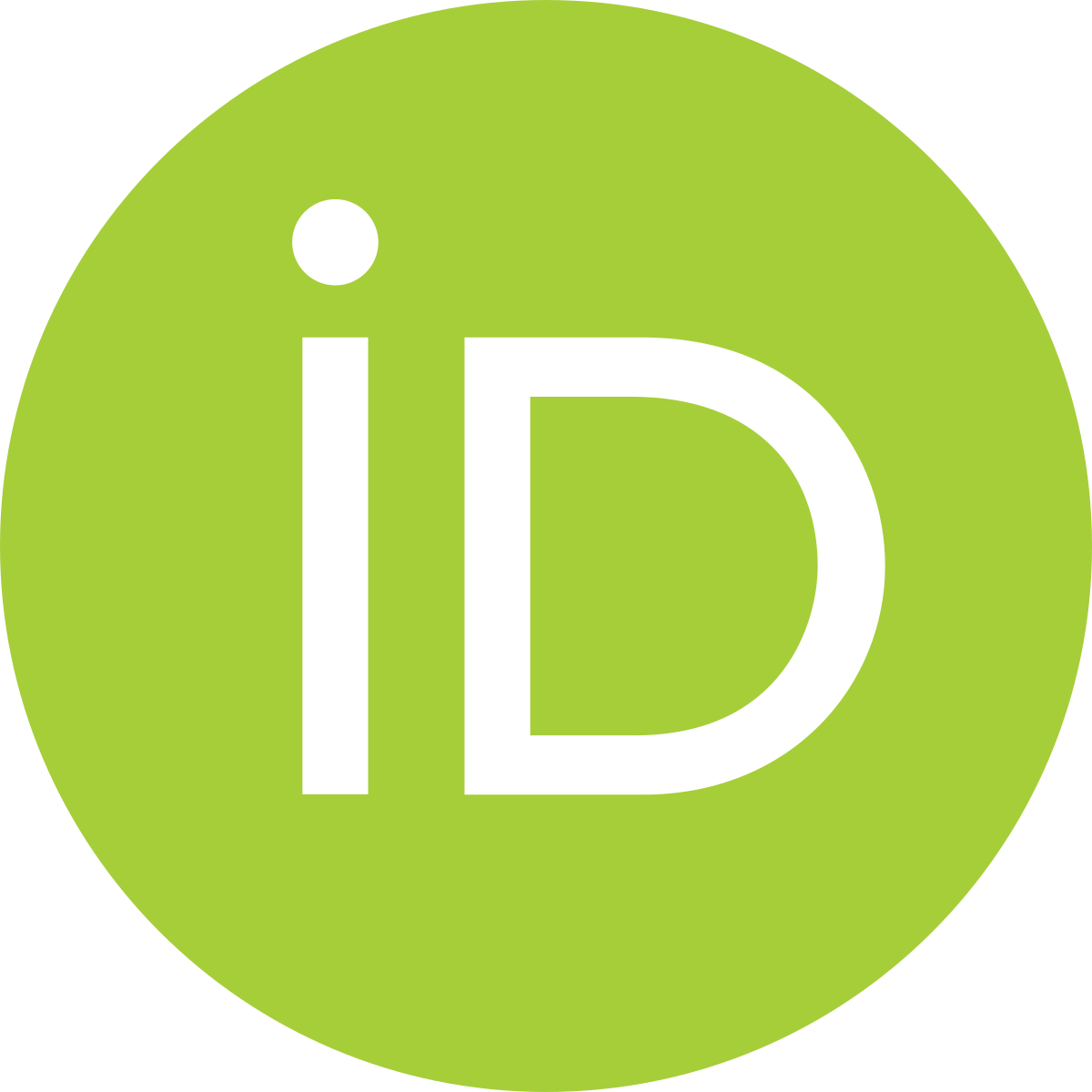}}}
\usepackage{amsmath}

\begin{document}
	
\title{Non-linear Transport in Non-centrosymmetric Systems: \\ From fundamentals to applications}
	
\author{Manuel Suárez Rodríguez\,\orcid{0000-0003-1186-2725}}
\affiliation{CIC nanoGUNE BRTA, 20018 Donostia-San Sebastián, Basque Country, Spain}
\affiliation{Department of Polymers and Advanced Materials: Physics, Chemistry and Technology UPV/EHU, 20018 Donostia--San Sebastián, Basque Country, Spain}
	
\author{Fernando De	Juan\,\orcid{0000-0001-6852-1484}}
\affiliation{Donostia International Physics Center, 20018 Donostia-San Sebastián, Basque Country, Spain}
\affiliation{IKERBASQUE, Basque Foundation for Science, 48009 Bilbao, Basque Country, Spain}

\author{Ivo Souza\,\orcid{0000-0001-9901-5058}}
\affiliation{Centro de Física de Materiales CSIC-UPV/EHU, 20018 Donostia-San Sebastián, Basque Country, Spain}
\affiliation{IKERBASQUE, Basque Foundation for Science, 48009 Bilbao, Basque Country, Spain}

\author{Marco Gobbi\,\orcid{0000-0002-4034-724X}}
\affiliation{Centro de Física de Materiales CSIC-UPV/EHU, 20018 Donostia-San Sebastián, Basque Country, Spain}
\affiliation{IKERBASQUE, Basque Foundation for Science, 48009 Bilbao, Basque Country, Spain}

\author{Fèlix Casanova\,\orcid{0000-0003-0316-2163}}
\affiliation{CIC nanoGUNE BRTA, 20018 Donostia-San Sebastián, Basque Country, Spain}
\affiliation{IKERBASQUE, Basque Foundation for Science, 48009 Bilbao, Basque Country, Spain}

\author{Luis E. Hueso\,\orcid{0000-0002-7918-8047}}
\affiliation{CIC nanoGUNE BRTA, 20018 Donostia-San Sebastián, Basque Country, Spain}
\affiliation{IKERBASQUE, Basque Foundation for Science, 48009 Bilbao, Basque Country, Spain}
	
\date{} % Leave this empty to suppress the default date

\makeatletter
\def\@date{\href{mailto:m.suarez@nanogune.eu}{m.suarez@nanogune.eu}, 
	\href{mailto:f.casanova@nanogune.eu}{f.casanova@nanogune.eu}, 
	\href{mailto:l.hueso@nanogune.eu}{l.hueso@nanogune.eu}}
\makeatother
	
\begin{abstract}
Ohm\textquotesingle s law has been a cornerstone of electronics since
its experimental discovery. This law establishes that in a conductive
system, the voltage is directly proportional to the current. Even when
time-reversal symmetry is disrupted, leading to the emergence of
magnetoresistance and Hall effects, the linear relationship between
voltage and current remains intact. However, recent experiments have
demonstrated a breakdown of Ohm's law in
non-centrosymmetric structures. In these systems, non-linear transport
effects are permitted with quadratic scaling between voltages and
currents. Here, we review the main demonstrations of non-linear
transport in non-centrosymmetric systems, analyzing the connection
between non-linear behavior and the system's symmetry.
Additionally, we delve into the microscopic mechanisms driving these
effects, such as Berry curvature dipole and Berry connection
polarizability. Finally, we highlight potential applications of
non-linear transport in spintronics and energy harvesting.
\end{abstract}
	
	\maketitle
	
\section{Introduction}

\begin{figure*}[t]
	\centering
	\includegraphics[width=1.0\linewidth]{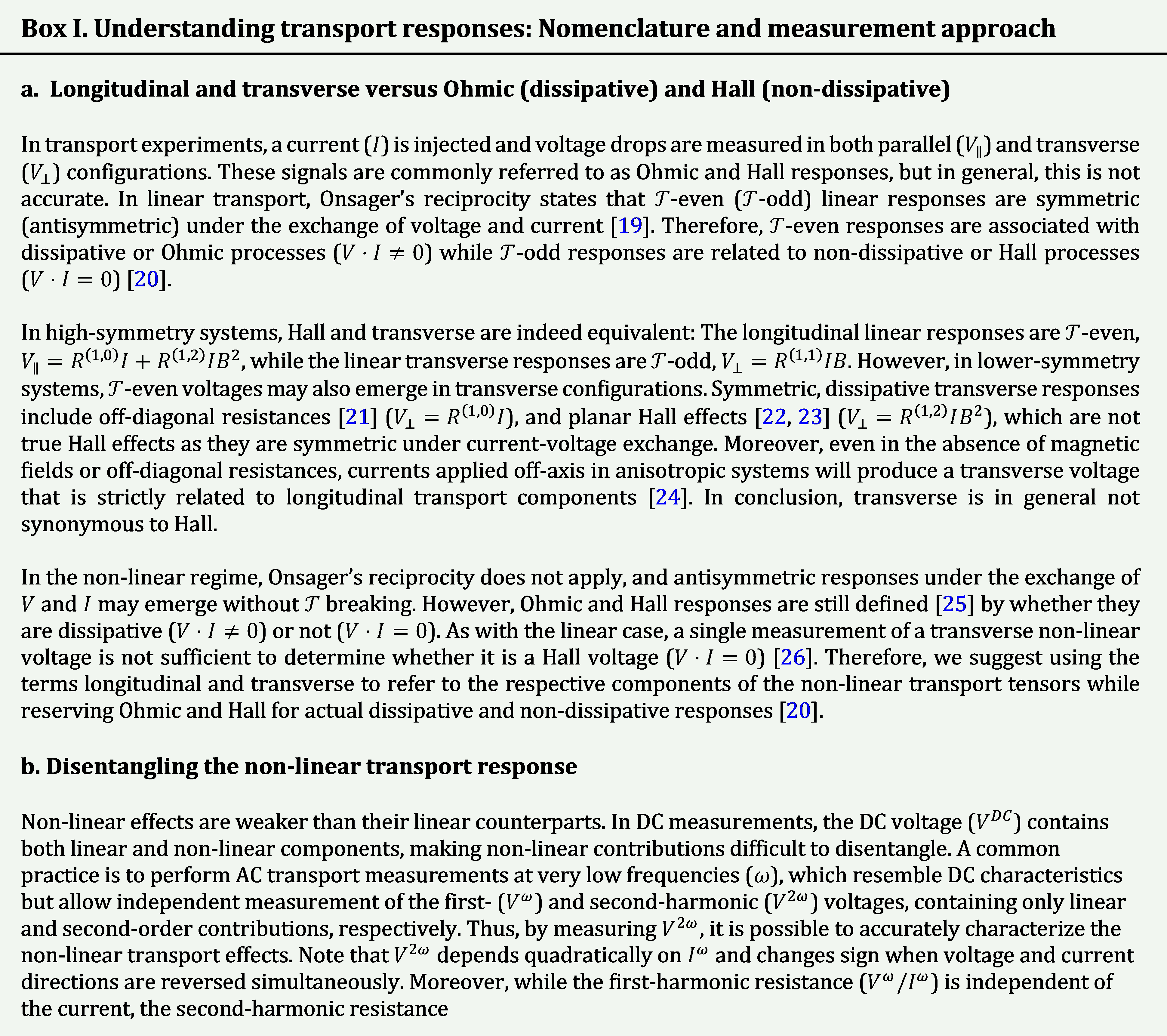}
	\label{box}
\end{figure*}

\begin{figure*}[t]
	\centering
	\includegraphics[width=0.88\linewidth]{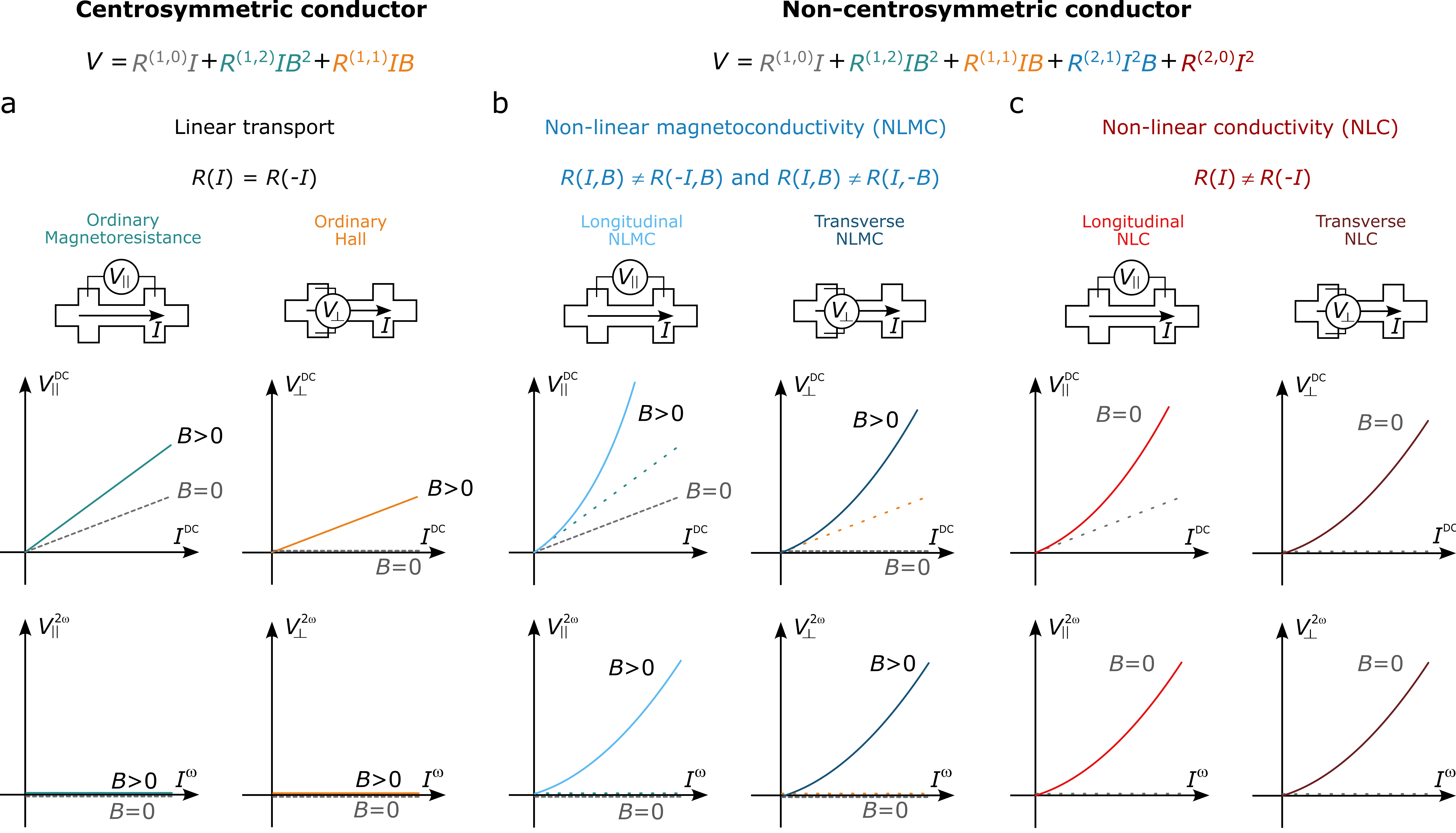}
	\caption{\textbf{Schematic of different transport components in
			centrosymmetric and non-centrosymmetric conductors.}
		\textbf{a}, Linear,
		\textbf{b}, NLMC and \textbf{c}, NLC are included, highlighting the
		corresponding measurement configurations (longitudinal, transverse).
		Additionally, the longitudinal and transverse DC
		(\(V_{\parallel}^{DC},V_{\bot}^{DC}\)) and second-harmonic
		(\(V_{\parallel}^{2\omega},\ V_{\bot}^{2\omega}\)) voltage versus
		current (\(I^{DC},\ I^{\omega})\) characteristics are included. When the
		overall signal (solid lines) is non-linear, the loosely dashed lines
		represent its linear contribution. In presence of an external magnetic
		field (\(B\)), the densely dashed lines refer to the signal at
		\(B = 0\). Importantly, in the second-harmonic voltage (bottom panels),
		only the non-linear contributions emerge. For clarity, we will
		refer to bilinear magnetoresistance/unidirectional magnetoresistance/electrical magnetochiral anisotropy and the non-linear
		planar Hall effect as longitudinal and transverse NLMC, respectively. We
		also note that the transverse NLC is often referred to as NLHE in the
		literature, but this term should be reserved exclusively for true Hall
		responses.}
	\label{fig:2.1}
\end{figure*}

In 1827, G. Ohm established that voltage (\(V\)) and current (\(I\)) are
linearly proportional, \(V = IR\), where \(R\) is the resistance of the
electrical conductor \cite{Ohm}. In his experiments, a
galvanometer was used to measure the current between the terminals of a
thermocouple, in which the voltage scales with the temperature gradient.
By experimenting with different materials, wire lengths, and temperature
gradients, he consistently observed that \(V\) and \(I\) are parallel
and linearly proportional. This empirical observation, known as Ohm's
law, remains one of the most important principles in modern electronics.
In 1856, W. Thomson discovered that the magnitude of the resistance can
be modified by the application of external magnetic
fields \cite{magnetoresistance}. Twenty-three years later, E. H. Hall revealed
that such magnetic fields, or the presence of magnetization, not only
affect the resistance value but also induce a voltage perpendicular to
the applied current \cite{hall} (Box \ref{box}.a). However, despite the
time-reversal symmetry (\(\mathcal{T}\)) breaking, the linear
relationship between voltage and current remains intact. In this case, we
can write in a more general fashion:
\(V = R^{(1,0)}I + R^{(1,2)}IB^{2} + R^{(1,1)}IB\), where \(R^{(1,0)}\),
\(R^{(1,2)}\) and \(R^{(1,1)}\) refer to the Ohmic, magnetoresistance
and Hall components, respectively (Fig. \ref{fig:2.1}.a). Along this manuscript we
use the notation \(R^{(a,b)}\), with \(a\) and \(b\) referring to the
current and magnetic field scaling order, respectively.

Non-centrosymmetric systems, where inversion symmetry (\(\mathcal{P}\))
is broken, do enable non-linear transport effects. In simple terms,
electrons experience different resistance when traveling in opposite
directions, as they are not equivalent. In this review, we focus on
second-order transport effects, though higher-order effects are also
possible \cite{13, 14, 15, 16}. As we will detail below, the
second-order effects cause the \(V - I\) characteristics of
\(\mathcal{P}\)-broken systems to be non-reciprocal, such that
\(V( + I) \neq - V( - I)\). In this regard, the \(V - I\)
characteristics of non-centrosymmetric materials resemble those of
classical diodes \cite{17}, while differing from other
systems, such as tunnel \cite{tunnel,19} or
Josephson \cite{20,21} junctions, that exhibit non-linear yet
reciprocal responses. To stress this peculiarity, the second-order
non-linear transport in non-centrosymmetric systems is often referred to
as non-reciprocal.  

Non-linear electrical transport effects in non-centrosymmetric systems
were first studied in chiral crystals. Pioneering studies had reported an
optical effect, referred to as magnetochiral anisotropy, which depends
on both the wave vector of the light and the external magnetic field
($B$)  \cite{22}. Shortly thereafter, its electrical
counterpart, electrical magnetochiral anisotropy, was also
observed  \cite{23,24}. In addition to the linear components,
the voltage has a contribution that scales quadratically with the
current and linearly with the external magnetic field (Fig. \ref{fig:2.1}.b). By
measuring the second-harmonic voltage (\(V^{2\omega}\)) in response to a
current at frequency \(\omega\) (\(I^{\omega}\)), it is possible to
isolate such contribution:
\(V^{2\omega} = R^{(2,1)}{(I^{\omega})}^{2}B\) (for details on the measurement technique, see Box \ref{box}.b). Subsequently, the effect was generalized to other non-centrosymmetric systems, such as
polar semiconductors  \cite{25}, topological
insulators  \cite{26} and two-dimensional electron
gases  \cite{27}. The non-linear transport effect in presence
of external magnetic fields may emerge in both longitudinal and
transverse configuration. \phantom{\cite{AHE, 5, offdiagonal, 108, PHE2, Manu4, 11, 12}} \hspace{-1.2cm} The longitudinal manifestation, where the
voltage is parallel to the current
(\(V_{\parallel}^{2\omega} = R^{(2,1)}{(I^{\omega})}^{2}B\)), has been
named bilinear magnetoelectric resistance  \cite{26,28},
bilinear magnetoresistance  \cite{29,30} or unidirectional
magnetoresistance  \cite{31,33}, since the second-harmonic
resistance (\(V^{2\omega}\text{/}I^{\omega}\)) depends linearly on both
the current and the magnetic field. The transverse effect, in which the
voltage is transverse to the current
(\(V_{\bot}^{2\omega} = R^{(2,1)}{(I^{\omega})}^{2}B\)), has also been
observed and referred to as the non-linear planar Hall
effect  \cite{34}, in analogy with its linear counterpart---the
planar Hall effect  \cite{PHE1,PHE2}. Additionally, other
terms such as non-reciprocal electrical transport  \cite{35,36}
can also be found in the literature, potentially leading to
misunderstandings among readers. In this review, we employ the term
``non-linear magnetoconductivity (NLMC)", for any \(V^{2\omega}\)
that scales quadratically with the current and linearly with the
magnetic field (\(V^{2\omega} = R^{(2,1)}{(I^{\omega})}^{2}B\)),
independently of the measurement configuration or system studied.

In the absence of an external magnetic field, non-linear transport
effects have also been recently observed. The non-linear Hall effect
(NLHE), which consists of a \(V^{2\omega}\) purely transverse to the
current, was first observed in WTe\textsubscript{2}  \cite{37,38}. It
immediately attracted much attention because, until that point, Hall
effects had been restricted to systems with broken \(\mathcal{T}\).
However, the Onsager reciprocity theorem does not apply in the
non-linear regime, allowing NLHE in systems with broken \(\mathcal{P}\)
but preserved \(\mathcal{T}\). Later, the longitudinal counterpart
(\(V_{\parallel}^{2\omega} = R^{(2,0)}{(I^{\omega})}^{2}\)), known as
the non-linear conductivity (NLC), was also
observed  \cite{39} (Fig. \ref{fig:2.1}.c). Both manifestations have been
intensively studied in polar systems  \cite{37,38}, two-dimensional electron gasses  \cite{69} and topological insulators  \cite{70}. Moreover, engineered structures have
also been developed to break \(\mathcal{P}\), yielding strong non-linear
responses  \cite{39, 40}. In general, the term Hall
should be reserved for non-dissipative responses, but it is often used
to describe any transverse voltage (for clarification on nomenclature,
see Box \ref{box}.a). A more precise approach is to use the term non-linear
conductivity (NLC) to broadly describe the non-linear transport effect
at zero field, with longitudinal and transverse NLC denoting the
respective components of the response. This definition leaves the term
NLHE for true Hall responses and standardizes the notation, allowing the
use of non-linear conductivity
(\(V^{2\omega} = R^{(2,0)}{(I^{\omega})}^{2}\)) and non-linear
magnetoconductivity (\(V^{2\omega} = R^{(2,1)}{(I^{\omega})}^{2}B\)) to
distinguish between non-linear transport effects in the absence and
presence of $B$.

In summary, in a centrosymmetric conductor, the voltage response to an
applied electrical current has three linear contributions with the
following dependence:
\(V = R^{(1,0)}I + R^{(1,2)}IB^{2} + R^{(1,1)}IB\). However, in a
non-centrosymmetric conductor, besides these three linear components,
two non-linear contributions emerge. Consequently, the voltage response
up to the second-order in the current follows the equation:
\begin{equation}
	V = R^{(1,0)}I + R^{(1,2)}IB^{2} + R^{(1,1)}IB + R^{(2,1)}I^{2}B + R^{(2,0)}I^{2},
	\label{eq:2.1}
\end{equation}
where \(R^{(2,1)}\) and \(R^{(2,0)}\) refer to the NLMC and the NLC
components, respectively (Fig. \ref{fig:2.1}). Although higher-order terms in the
magnetic field are theoretically possible, their expected lower
magnitude renders experimental detection challenging. Previous reviews
have discussed general non-linear phenomena \cite{42, 43, 44} or
did focus on transport effects, but only in specific systems, such as
van der Waals nanostructures \cite{45}, or on a particular
effect, such as NLHE \cite{46, 47, 48}. However, despite the rapid
growth of the field, a comprehensive review offering a general yet
thorough overview of non-linear transport phenomena in
non-centrosymmetric systems, while incorporating the latest
contributions, is still lacking. In this article, the main reports on
non-linear transport effects in non-centrosymmetric systems will be
reviewed, covering polar and chiral systems, as well as magnetic
materials and engineered structures. Importantly, given the intimate
link between non-linear effects and material symmetry, the procedure to
predict both NLMC and NLC from the point group symmetry of a system will
be explained. Moreover, one section will be devoted to discussing the
physical microscopic mechanisms behind these non-linear effects,
outlining strategies to disentangle them in experiments. Finally,
different applications of non-linear transport effects will be explored,
with a particular focus on spintronics and energy harvesting.
	
\section{Non-centrosymmetric systems} \label{sec:2.1}

Systems in which inversion symmetry (\(\mathcal{P}\)) is broken are
called non-centrosymmetric. The inversion symmetry operation involves
transforming a coordinate \((x,\ y,\ z)\) to \(( - x,\  - y, - z)\).
Therefore, in non-centrosymmetric systems, moving in one direction is
not equivalent to moving in the opposite direction. This is exactly what
happens when an electrical current is applied: electrons traveling in
one direction experience different resistance than those traveling in
the opposite direction, leading to non-linear transport effects. In this
section, we will review the main reports on non-linear transport effects
in non-centrosymmetric systems, both in the presence and absence of
external magnetic fields. Interestingly, the non-linear responses across
all configurations can be predicted using the point group of the system. Thus, we will begin by discussing this analysis, which will provide a clearer understanding of the experimental results presented
later.
	
\subsection{Crystal symmetry origin} \label{sec:2.1.1}

Non-linear transport effects are intrinsically linked to the symmetry of
the system under study. As a result, the material\textquotesingle s
point group symmetry can be used to predict its non-linear response. In
centrosymmetric point groups, non-linear transport is forbidden.
However, once \(\mathcal{P}\) is broken, non-linear transport can occur
in specific configurations. The number and nature of these
configurations depend on which additional symmetry elements are broken.
While measured voltages in response to applied currents are the
magnitudes used experimentally, it is more convenient to use current
densities (\(j\)) in response to applied electric fields (\(E\)) for the
theoretical description of the phenomena. Below, we present the
configurations of \(j\), \(E\) and \(B\) for which the non-linear
transport is allowed in each point group (Table \ref{tab:2.1}) and discuss the
procedure for linking them to the experimental results.

\begin{table*}[t]
	\centering
	\caption{\textbf{Summary of the different non-linear transport components allowed in each non-magnetic crystalline point group. }
		In the notation “$ij$”, the letter “$i$” refers to the direction of the second-harmonic current density ($j_i^{2\omega}$) in response to electric fields along “$j$” ($E_{j}^{\omega}E_{j}^{\omega}$). Forbidden NLC components are marked with a red cross, whereas allowed ones are marked with a green tick. For the NLMC, forbidden components are also tagged with a red cross, whereas allowed ones list, in the corresponding box, the required directions of $B$ to induce the effect. The non-centrosymmetric point groups are classified into 4 types: polar (magenta), chiral (orange), both (blue), or neither (green). The centrosymmetric point groups are shaded in blue.}
	\includegraphics[width=0.87\linewidth]{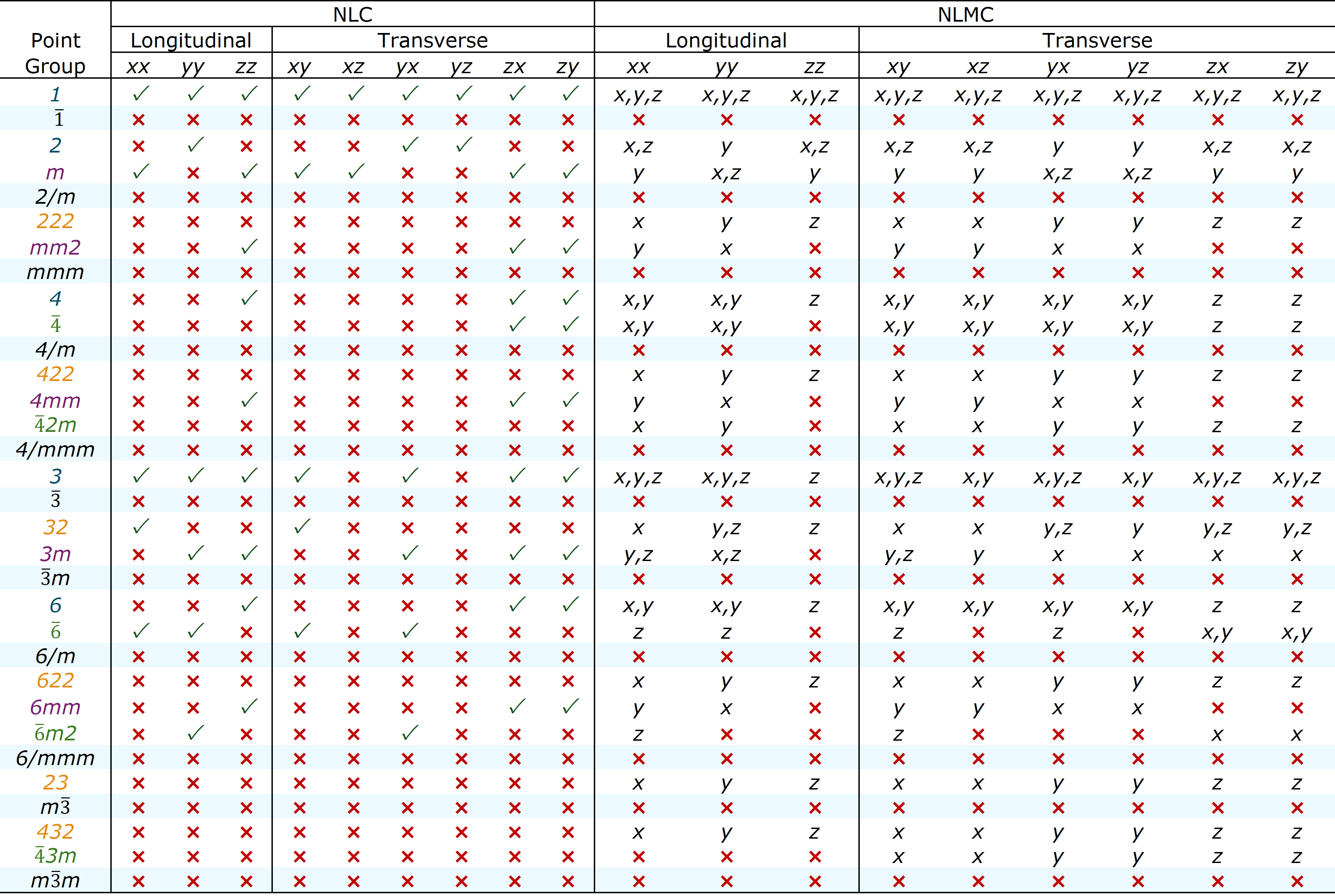}
	\label{tab:2.1}
\end{table*}

The second-order non-linear current density\hspace{0pt}
(\(j_{i}^{2\omega}\)) in response to an electric field
(\(E_{j,k}^{\omega}\))\hspace{0pt} can be expressed through the
material\textquotesingle s non-linear conductivity tensor
(\(\sigma^{(2,0)}_{ijk}\)) \hspace{0pt}as:
\(j_{i}^{2\omega} = \sigma^{(2,0)}_{ijk}E_{j}^{\omega}E_{k}^{\omega}\)\hspace{0pt}.
This tensor is known for each point group and can be found in the Bilbao
Crystallographic Server under the name of non-linear
susceptibility  \cite{bilbao}. Although some physical mechanisms
are exclusive to the Hall response (see section \ref{sec:2.2}), the tensor
incorporates both Ohmic and Hall responses  \cite{11}, along
with longitudinal and transverse components, underscoring
the suitability of using the common term NLC. As mentioned above,
although the theoretical description is derived in terms of non-linear
conductivities (\(\sigma^{(2,0)}_{ijk}\)), experimentally, it is useful to express
the equation in terms of the non-linear resistivity
(\(\rho_{ijk}^{(2,0)}\)),
\(E_{i}^{2\omega} = \rho_{ijk}^{(2,0)}j_{j}^{\omega}j_{k}^{\omega}\).
By inverting the matrix (see appendix \hyperref[s:note]{A}), the relationship between non-linear resistivity
and conductivity is obtained:
\hspace{0pt}\(\rho_{ijk}^{(2,0)}\)=\(- \rho^{(1,0)}_{il}\sigma^{(2,0)}_{lmn}\rho^{(1,0)}_{mj}\rho^{(1,0)}_{nk}\),
where \(\rho^{(1,0)}_{ij}\) refer to the Ohmic resistivity
(\(E_{i}^{\omega} = \rho^{(1,0)}_{ij}j_{j}^{\omega}\)). Therefore, the
second-order electric field (\(E_{i}^{2\omega}\)) in response to a
current density (\(j_{j,k}^{\omega}\)) can be written as:
\(E_{i}^{2\omega} = - \rho^{(1,0)}_{il}\sigma^{(2,0)}_{lmn}\rho^{(1,0)}_{mj}\rho^{(1,0)}_{nk}j_{j}^{\omega}j_{k}^{\omega}\). Note that non-centrosymmetric systems
generally exhibit anisotropic resistivity. Thus, as in the linear regime
(see Box \ref{box}.a), when currents are applied off-axis, non-linear
longitudinal tensor components may emerge in a transverse configuration,
and vice versa. Moreover, in triclinic and
monoclinic systems (\(\rho^{(1,0)}_{ij} \neq 0\) with \(i \neq j\)), even when
voltages are measured in response to currents along the principal axes,
both longitudinal and transverse components may arise in both
measurement configurations.

For effects induced by a magnetic field, the same procedure is followed.
The second-order non-linear current density (\(j_{i}^{2\omega}\)) in
response to an electric field (\(E_{j,k}^{\omega}\))\hspace{0pt} and a
magnetic field (\(B_{l}\)) can be written in terms of the NLMC tensor
(\(\sigma_{ijkl}^{(2,1)}\)) as:
\(j_{i}^{2\omega} = \sigma_{ijkl}^{(2,1)}E_{j}^{\omega}E_{k}^{\omega}B_{l}\).
The NLMC tensor is not directly available in the Bilbao Crystallographic
Server database but can be generated using the Jahn symbol
\(eV\lbrack V2\rbrack V\)  \cite{51}. As in the previous case, it is
convenient to write the equation in terms of the non-linear
magnetoresistivity (\(\rho_{ijkl}^{(2,1)}\)),
\(E_{i}^{2\omega} = \rho^{(2,1)}_{ijkl}j_{j}^{\omega}j_{k}^{\omega}B_{l}\).
By inverting the matrix (see appendix \hyperref[s:note]{A}), the NLMC contribution to the second-order
electric field (\(E_{i}^{2\omega}\)) in response to a current density
(\(j_{j,k}^{\omega}\)) and a magnetic field (\(B_{l}\)) is obtained:
\(E_{i}^{2\omega} = - \rho_{im}^{(1,0)}\sigma_{mnql}^{(2,1)}\rho^{(1,0)}_{nj}\rho^{(1,0)}_{qk}j_{j}^{\omega}j_{k}^{\omega}B_{l}\).
Moreover, an additional contribution arising from a combination of
the non-linear conductivity and the ordinary Hall effect is also
possible (for more details, see appendix \hyperref[s:note]{A}).

In Table \ref{tab:2.1}, we summarize the \(j\), \(E\) and \(B\) configurations for
which the non-linear transport is allowed in each non-magnetic point
group, both in the presence and absence of external magnetic fields. For
the NLMC, the required direction of \(B\) is also indicated. Extra
components may be allowed when \(E_{j}^{\omega}\) is applied between two
principal axes
(\(j_{i}^{2\omega} \propto E_{j}^{\omega}E_{k}^{\omega}\), with
\(j \neq k\)), but we focus on the cases when \(E_{j}^{\omega}\) is
along a principal axis because they are the most studied configurations
in experiments. Therefore, by referring to Table \ref{tab:2.1}, it is possible to
directly predict the non-linear components in any non-centrosymmetric
systems with a known point group. Additionally, this enables
verification of whether the observed non-linear signals align with the
material's symmetry. If they do not, surface
contributions, thermal effects, or other artifacts may be experimentally
significant.

\subsection{Experimental observations} \label{sec:2.1.2}

\emph{\textbf{B.1. Polar systems.} --} These systems exhibit structural
asymmetry characterized by a well-defined directionality, which includes
two or more points that remain unmoved under every symmetry operation.
Depending on the number of unmoved points, they can define a line, a
plane, or all of space (polar direction). These systems usually host
electric polarization, which must lie parallel to the polar direction.
Spontaneous and modulable electric polarizations result in ferroelectric
materials. Interestingly, the modulation of electric polarization
affects the symmetry of the material and, therefore, it will have an
impact on the non-linear transport.

\begin{figure*}[t]
	\centering
	\includegraphics[width=0.94\linewidth]{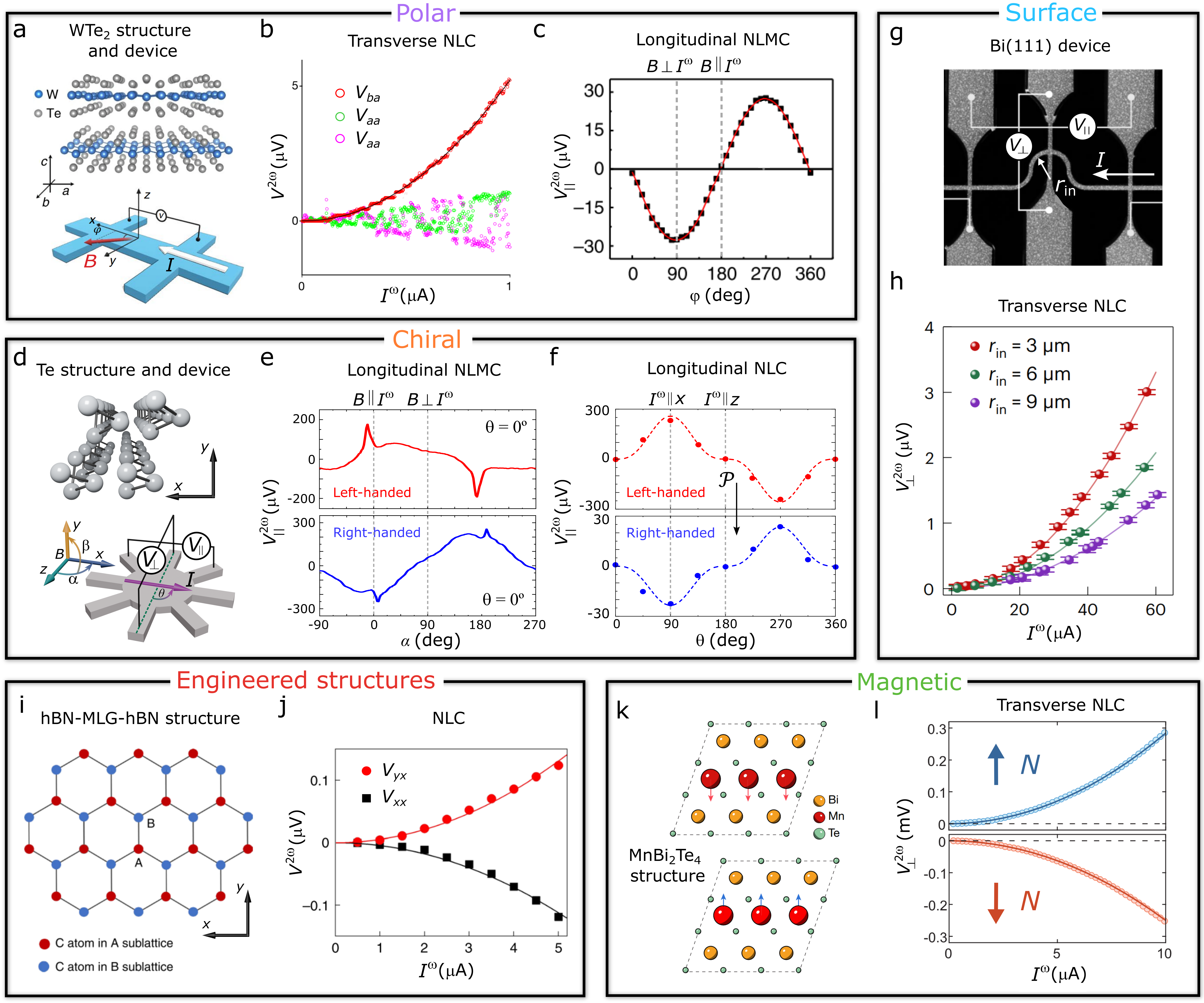}
	\caption{\textbf{Non-linear transport in non-centrosymmetric
			systems.}
		\textbf{a}, WTe\textsubscript{2} structure and geometry of device used in \textbf{c}. \textbf{b},
		Second-harmonic voltage (\(V^{2\omega}\)) as a function of current (\(I^{\omega}\)) in WTe\textsubscript{2}.
		The non-linear signal is only observed in transverse configuration.
		\textbf{c}, \(V^{2\omega}\) as a function of the in-plane
		\(\varphi\)-angle between an external magnetic field (\(B\)) and
		\(I^{\omega}\). The signal is maximum when \(B\) and \(I^{\omega}\) are
		mutually perpendicular, in agreement with WTe\textsubscript{2} polar
		symmetry. \(V^{2\omega}\) is detected parallel to \(I^{\omega}\).
		\textbf{d}, Te structure and geometry of
		device used in \textbf{e} and \textbf{f}. \textbf{e}, \(V^{2\omega}\) as
		a function of the in-plane \(\alpha\)-angle between \(B\) and
		\(I^{\omega}\) along the chiral $z$-axis. The maximum signal is
		recorded when $B$ and \(I^{\omega}\) are parallel, in agreement
		with Te chiral symmetry. \textbf{f}, \(V^{2\omega}\) as a function of
		the \(\theta\)-angle between \(I^{\omega}\) and the chiral
		$z$-axis. The signal is maximum when \(I^{\omega}\) is along the
		$x$-axis, in agreement with Te symmetry. In \textbf{e} and
		\textbf{f}, \(V^{2\omega}\) is detected in parallel configuration and
		changes sign between left- and right-handed Te crystals. \textbf{g},
		Bi(111) device geometry. \textbf{h}, Transverse \(V^{2\omega}\) as a function of \(I^{\omega}\) in Bi (111) at room temperature. The signal is enhanced by arc-shaped geometries. \textbf{i}, Inequivalent carbon sublattices in hBN-MLG-hBN heterostructures.
		\textbf{j}, \(V^{2\omega}\) as a function of
		\(I^{\omega}\) in hBN-MLG-hBN. Non-linear signals are recorded in both
		longitudinal and transverse configurations. \textbf{k},
		MnBi\textsubscript{2}Te\textsubscript{4} structure. \textbf{l},
		Transverse \(V^{2\omega}\) as a function of \(I^{\omega}\) in
		MnBi\textsubscript{2}Te\textsubscript{4}. The non-linear signal changes
		sign by reversing the Néel vector. Figures adapted with permission from: \textbf{a},\textbf{c}, Ref. \cite{58} (CC-BY-4.0); \textbf{b}, Ref. \cite{37};  \textbf{g}-\textbf{h}, Ref. \cite{74}; \textbf{i}-\textbf{j}, Ref. \cite{39}; \textbf{k}, Ref. \cite{90} (Springer Nature Limited); and \textbf{l}, Ref. \cite{12} (AAAS).}
	\label{fig:2.2}
\end{figure*}

In polar WTe\textsubscript{2} (Fig. \ref{fig:2.2}.a), transverse NLC was observed and associated with
the purely transverse Berry curvature dipole (BCD) mechanism, marking
the first report of a Hall effect in a time-reversal-invariant system
and quickly attracting significant attention  \cite{37,38} (Fig. \ref{fig:2.2}.b). However, extrinsic mechanisms were also found to be
important  \cite{38}. Shortly after, also in WTe\textsubscript{2},
ferroelectric control of the non-linear signal was demonstrated, showing
that the NLC can be exploited to detect the electric polarization
orientation  \cite{52}. The NLC has been reported in other
polar transition metal dichalcogenides (TMDCs), such as
MoTe\textsubscript{2}  \cite{53}, and in out-of-plane configurations, with
voltage recorded perpendicular to the TMDC plane  \cite{54}.
Importantly, the effect was observed in chemical vapor deposition-grown
samples, highlighting its potential for real-world
applications  \cite{53}. Non-TMDC polar materials have also
shown NLC with different microscopic mechanisms. For instance, in
NbIrTe\textsubscript{4}  \cite{55}, it is dominated by BCD,
whereas in BiTeBr  \cite{56} and BaMnSb\textsubscript{2}  \cite{57},
extrinsic mechanisms are the major contribution.

In WTe\textsubscript{2}, NLMC was reported as well, in both
longitudinal \cite{28,58,59} (Fig. \ref{fig:2.2}.c) and
transverse  \cite{34,59} configurations. The
crystallographic direction dependence was
studied  \cite{28,58}, yielding results compatible with its
polar symmetry. Indeed, the NLMC signal is consistently
observed when the magnetic field is perpendicular to the voltage
detection direction. Similar results were obtained in the polar bulk
semiconductor BiTeBr  \cite{25}. The symmetry of the NLMC was associated with the helical spin texture of these systems,
showing that it can be leveraged for spin texture detection (see section \ref{sec:2.3.1}).

\emph{\textbf{B.2. Chiral systems.} --} The term \emph{chiral} refers to an
object that produces a non-superimposable mirror image of itself, with
human hands being the classical example. More precisely, a chiral system
lacks any symmetry operation that reverses orientation (mirror
reflection, inversion, or rotoinversion) and therefore cannot be mapped
onto its mirror image through rotations and translations alone. Most of
the recent experimental reports on non-linear transport in chiral
materials have focused on elemental tellurium (Te). This crystal is
formed by covalently bonded Te helices, stacked together by van der
Waals interactions (Fig. \ref{fig:2.2}.d). Furthermore, it is a naturally
\emph{p}-doped semiconductor that presents good electronic transport
properties, and its handedness can be identified through transmission
electron microscopy \cite{60}.

\vspace{0.2cm}

After a first report in artificially made bismuth
helices \cite{23}, NLMC was demonstrated in a pristine
system---macroscopic Te crystals \cite{24}. However, the
observed signal does not align with Te chiral symmetry. Later studies
using Te flakes revealed the predicted behavior \cite{Manu1},
highlighting the critical role of symmetry analysis in ruling out
potential artifacts (Table S1). Additionally, gate tuneability was
shown, and a sign change between right- and left-handed flakes was
detected \cite{Manu1} (Fig. \ref{fig:2.2}.e). These findings were confirmed in
subsequent experiments \cite{61}. Recently, the
crystallographic direction dependence of NLMC was studied in both
parallel and transverse configurations, perfectly matching the expected
symmetry for Te \cite{Manu4}. In contrast to polar systems, NLMC
is observed when the magnetic field is parallel to the voltage detection
direction, aligning with the radial spin texture of chiral
systems \cite{Manu1} (for more details, see section \ref{sec:2.3.1}). In
chiral helimagnetic CrNb\textsubscript{3}S\textsubscript{6}, an NLMC
signal was reported, which is enhanced below the magnetic ordering
temperature, illustrating the influence of chiral magnetic order on the
non-linear transport \cite{63}. Additionally, NLMC was also
observed in a chiral molecular conductor, confirming the chiral
character of charge transport in molecular
materials \cite{64}.

In the absence of external magnetic fields, longitudinal NLC was
reported in Te and was demonstrated to be odd under spatial
inversion \cite{Manu2} (Fig. \ref{fig:2.2}.f). In agreement with the NLMC
measurements \cite{Manu4}, the resistivity dependence of the
non-linear signal suggests that extrinsic scattering from dynamic
sources is the dominant microscopic mechanism \cite{Manu2}. B.
Cheng \emph{et al.} reported transverse NLC effect in
Te \cite{65}; however, the observed signal does not align with
Te symmetry (Table \ref{tab:2.1}, point group \emph{32}). The authors propose that
surface symmetry breaking explains the discrepancy \cite{65}.

\emph{\textbf{B.3. Surfaces and interfaces.} --} In non-centrosymmetric
crystals, \(\mathcal{P}\) is broken in the bulk. However, all systems
break \(\mathcal{P}\) at the surface level. Therefore, if electronic
transport is dominated by surface states, such as in two-dimensional
electron gases (2DEGs) or topological insulators (TIs), non-linear
phenomena may arise even when \(\mathcal{P}\) is preserved in the
bulk \cite{66}.

The NLMC has been widely observed in 2DEGs \cite{27,29,31,33}
and TIs \cite{26,67,68}. In these systems, \(\mathcal{P}\) is
broken at the surface level, while rotation axis and mirror planes
around the surface normal direction are generally preserved. Under these
conditions, and similar to polar systems, helical spin textures emerge
and NLMC signals are recorded with in-plane magnetic fields
perpendicular to the current \cite{29,31,67} (for more
details, see section \ref{sec:2.3.1}). Interestingly, when rotation axes and mirror
planes perpendicular to the surface are also broken, NLMC signals with
out-of-plane magnetic fields are observed as well \cite{26,27}. In general, the NLMC has been measured in longitudinal
configuration; however, it has also been reported in transverse
configuration in both TIs and 2DEGs \cite{34}. Furthermore, in
these systems, the ratio between surface and bulk transport can be
controlled by electrostatic gating, leading to highly modulable NLMC
signals \cite{29,31,67}.

\begin{table*}[t]
	\centering
	\caption{\textbf{Summary of the main experimental results on non-linear transport in non-centrosymmetric materials.}
		Non-linear effect (NLC, NLMC, anomalous NLMC), material and measurement configuration (Conf.) are detailed. Second-harmonic output voltage ($V^{2\omega}$), applied current ($I^{\omega}$), temperature ($T$) range, linear resistivity ($\rho$) and magnetic field ($B$) (for NLMC) are included. The temperature at which the other parameters are reported is given in parenthesis. From these parameters, the non-linear conductivity ($\sigma^{(2,0)}$) and resistivity ($\rho^{(2,0)}$) in the absence of $B$, and the non-linear magnetoconductivity ($\sigma^{(2,1)}$) and magnetoresistivity ($\rho^{(2,1)}$) were calculated. We note that some values have been extracted from figures in the references and are, therefore, approximate.}
	\includegraphics[width=0.872\linewidth]{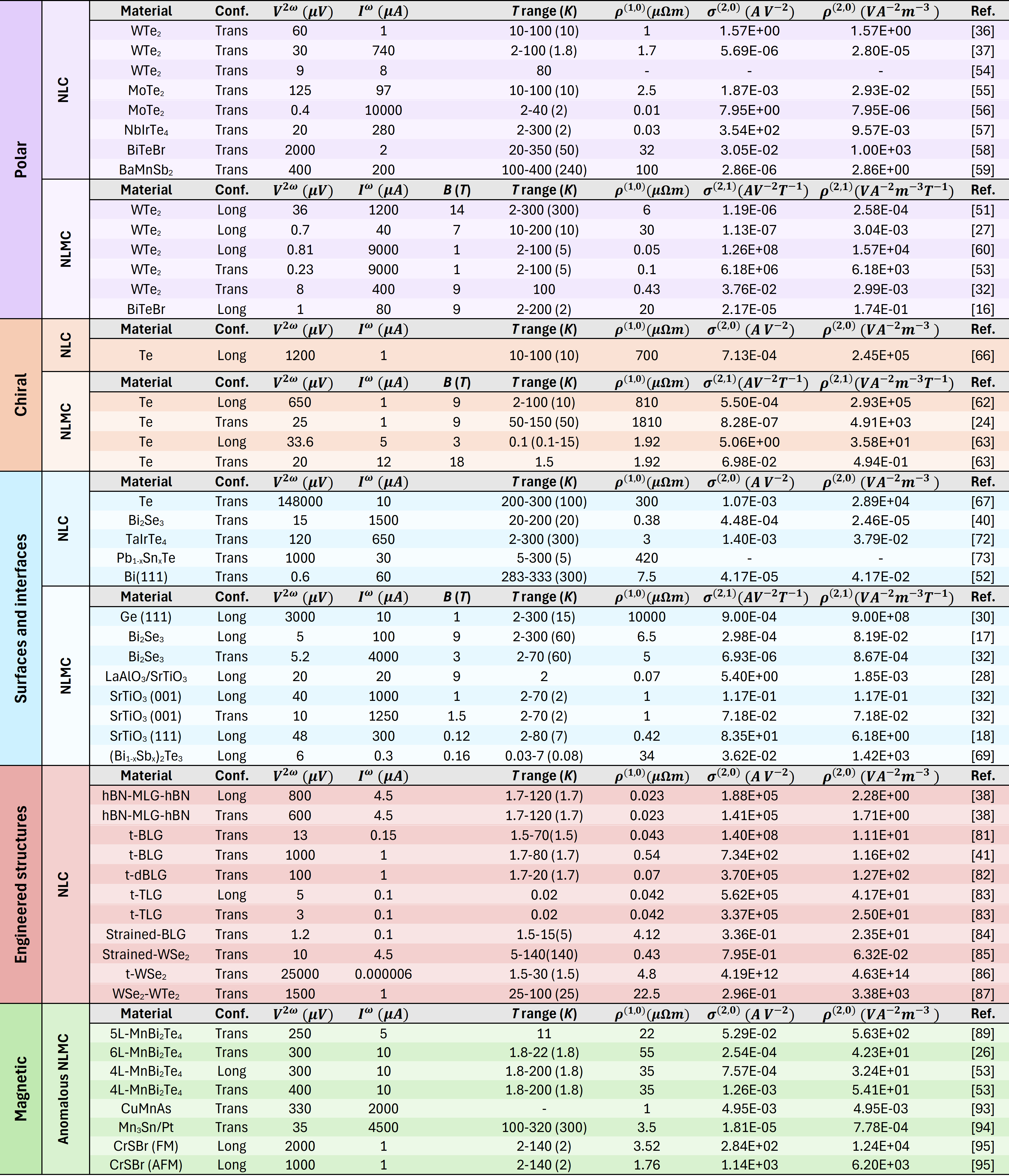}
	\label{tab:2.2}
\end{table*}

NLC has been reported in 2DEGs \cite{69},
TIs \cite{70} and Weyl semimetals \cite{71},
surviving up to room temperature \cite{72}. Moreover,
non-volatile ferroic switching of the non-linear signal has been
demonstrated in
Pb\textsubscript{1--\emph{x}}Sn\emph{\textsubscript{x}}Te \cite{73}. Importantly, NLC was observed at room temperature
on the (111) surface of bismuth (Bi), which is easily grown and
compatible with CMOS technology and flexible
substrates \cite{74} (Fig. \ref{fig:2.2}.g). The reported non-linear signal, which is
connected to the side-jump and skew-scattering mechanisms, can be
enhanced with arc-shaped geometries \cite{75}, offering a new
strategy to manipulate non-linear phenomena (Fig. \ref{fig:2.2}.h).

\emph{\textbf{B.4. Engineered heterostructures.} --}Crystalline systems found in nature or grown by standard methods are often centrosymmetric, which restricts the number of materials capable of hosting non-linear transport effects. However, material engineering approaches, such as stacking and twisting different layers \cite{jacobo}, show promise in overcoming this limitation. In this regard, van der Waals systems, which can be easily exfoliated into single or multiple atomic layers and subsequently stretched or combined, stand out as an ideal platform \cite{76, carmine2, 77}. 

Graphene was the first van der Waals material to be exfoliated down to a
single layer, opening a new era in material science \cite{78}.
However, single layer graphene (SLG) presents a centrosymmetric
structure, which forbids non-linear transport effects. Importantly,
graphene is generally capped or sandwiched by hexagonal boron nitride
(hBN) to improve its electronic properties. The presence of hBN breaks
\(\mathcal{P}\), enabling non-linear transport phenomena (Fig. \ref{fig:2.2}.i). The
hBN/SLG/hBN heterostructure was the first system to demonstrate NLC in
both longitudinal and transverse configurations within a single
structure \cite{39} (Fig. \ref{fig:2.2}.j). The longitudinal response
cannot originate from the Berry curvature dipole (BCD), with skew
scattering identified as the dominant microscopic mechanism. Recently,
it was discovered that breaking \(\mathcal{P}\) by twisting graphene
layers provides unprecedented possibilities to introduce completely new
physics, such as superconductivity \cite{79}. In the field of
non-linear transport, twisted bilayer graphene
(t-BLG) \cite{40,80}, twisted double bilayer graphene
(t-dBLG) \cite{81}, and twisted trilayer graphene
(t-TLG) \cite{82} have all reported NLC signals with giant
outputs. In the absence of hBN or twisting, bilayer graphene can be
artificially corrugated by a lithographically patterned strain, breaking
\(\mathcal{P}\) and exhibiting NLC signals \cite{83}.

Other van der Waals systems, different from graphene, have also been
engineered to break the required symmetries. For instance,
stretching \cite{84} or twisting \cite{85}
WSe\textsubscript{2} results in surprisingly high NLC signals. Moreover,
a heterostructure formed by WSe\textsubscript{2} and
WTe\textsubscript{2} has demonstrated ferroelectric control of BCD,
giving rise to highly modulable non-linear
phenomena \cite{86}. Other approaches, such as applying an
additional electric field to generate a sizable BCD \cite{87},
have also been reported.

\emph{\textbf{B.5. Magnetic systems.} --}In non-centrosymmetric magnetic
materials, besides \(\mathcal{P}\), \(\mathcal{T}\) is also broken.
Therefore, even though they could be classified under previous sections,
we discuss them separately. Indeed, in this case non-linear effects
governed by the magnetization (\(M\)) or Néel vector (\(N\)) emerge.
These additional contributions, where \(M\) or \(N\) play a role similar
to \(B\), can be interpreted as a NLC dependent on the magnetic vectors,
but also as an anomalous version of NLMC, highlighting their distinct
origin and \(\mathcal{T}\)-odd nature (\(V^{2\omega} \propto I^{2}M\) or
\(V^{2\omega} \propto I^{2}N\)).

MnBi\textsubscript{2}Te\textsubscript{4} is a non-centrosymmetric topological insulator in which the Mn
spins are aligned parallel in each layer but are coupled antiparallel to
adjacent layers (Fig. \ref{fig:2.2}.k). Therefore, odd-layer MnBi\textsubscript{2}Te\textsubscript{4} is a
topological ferromagnet, whereas even-layer MnBi\textsubscript{2}Te\textsubscript{4} is a topological
antiferromagnet. On the one hand, odd-layer MnBi\textsubscript{2}Te\textsubscript{4} presents chiral
edge states, which lead to non-linear output voltages with different
signs at opposite edges \cite{88}. Importantly, by reversing
the magnetization direction, the sign of the non-linear output also
changes. Similar results have been found in Cr-doped
(Bi,Sb)\textsubscript{2}Te\textsubscript{3} \cite{89}. On the other hand, even-layer MnBi\textsubscript{2}Te\textsubscript{4} has been reported to exhibit
non-linear transport in both transverse \cite{12,90} and
longitudinal \cite{90,91} configuration. In these systems,
both \(\mathcal{P}\) and \(\mathcal{T}\) symmetries are broken, but
\(\mathcal{PT}\) symmetry is preserved below the Néel temperature.
Importantly, the emergence of \(\mathcal{PT}\) symmetry cancels the BCD,
so it cannot be the origin of the reported effects. The field-induced
correction to Berry curvature, also called Berry connection
polarizability (BCP), or positional shift correction, was proposed to
give rise to a \(\mathcal{PT}\) invariant non-linear transport
effect \cite{92} and is believed to be the main mechanism
behind these experiments (for more details, see section \ref{sec:2.2.3}). This mechanism is often quoted to be related to a band-normalized quantum metric, and has gained much attention in the community. Remarkably, the reported signal can be modulated by reversing the Néel vector (Fig. \ref{fig:2.2}.l). However,
it is important to note that the control of the non-linear signal
through Néel vector manipulation had previously been demonstrated in
CuMnAs using current pulses \cite{93}. More recently,
manipulation with moderate magnetic fields has also been reported in
Mn\textsubscript{3}Sn, leading to highly tunable non-linear
phenomena \cite{94}. Even-layer CrSBr is a non-centrosymmetric
antiferromagnet at low fields that, with the application of moderate
magnetic fields, undergoes a metamagnetic transition to a ferromagnetic
state. In our lab, we have shown anomalous NLMC in CrSBr, which is fully
modulable by both magnetization and Néel vector \cite{95}. All
these results demonstrate that the anomalous NLMC can be used to
determine the orientation of the magnetization in ferromagnets and Néel
vector in antiferromagnets, opening the door to the development of
electrical readout in magnetic devices based on non-linear transport
effects.

Finally, it is important to note that some systems possess a centrosymmetric crystal structure but undergo $\mathcal{P}$-breaking transitions, such as charge density wave, topological, and magnetic phase transitions, thereby enabling non-linear transport effects. Consequently, non-linear transport measurements can be exploited to probe such transitions \cite{errea, 96, 97, 98}.
	
\section{Physical mechanisms} \label{sec:2.2}

Beyond their significance from a material-symmetry perspective, non-linear effects have garnered considerable attention in quantum condensed matter. These effects enable the experimental observation of previously hidden physical mechanisms, leading to the discovery and extensive study of novel quantum properties such as the Berry curvature dipole and Berry connection polarizability. Although some physical mechanisms contribute to non-linear effects both with and without $B$, we will address them separately for clarity. Indeed, in the absence of $B$, the $\mathcal{T}$-even responses scale with odd powers of the scattering time ($\tau$), while the $\mathcal{T}$-odd responses scale with even powers. Conversely, in the presence of $B$, the $\mathcal{T}$-even responses depend on even powers of $\tau$, while the $\mathcal{T}$-odd responses depend on odd powers. This scaling will be important for disentangling the different mechanisms in experiments. Additionally, in non-centrosymmetric magnetic systems, $\mathcal{T}$ is broken even in the absence of $B$, leading to unique physical mechanisms dependent on the magnetic vectors. Therefore, we dedicate a section to this additional non-linear transport contribution, which can be regarded as anomalous NLMC (for more details, see section \ref{sec:2.1.2}.5).

\subsection{Non-linear conductivity} \label{sec:2.2.1}

The NLC, particularly its non-dissipate manifestation known as the NLHE, has led to a new frontier in electronics. Linear Hall effects are forbidden in systems with time-reversal symmetry ($\mathcal{T}$) due to Onsager’s reciprocity (Box \ref{box}.a). However, in systems where $\mathcal{P}$ is broken, Hall effects are allowed in the non-linear regime. As a result, it is possible to measure a non-linear, non-dissipative transverse voltage under $\mathcal{T}$ conditions. This effect, when first reported experimentally, was linked to an intrinsic physical property known as the Berry curvature dipole. However, shortly afterward, subsequent experiments emphasized the significant role of extrinsic mechanisms in non-linear transport, which also contribute to dissipative processes and in a longitudinal configuration. In this section, we discuss both intrinsic and extrinsic mechanisms and explain how to experimentally disentangle them.

\emph{\textbf{A.1. Berry curvature dipole.} --} The intrinsic contribution to
the conventional linear Hall effect is the integral of the Berry
curvature (\(\Omega\)) over the Fermi sea. In a
\(\mathcal{T}\)-invariant system, the Berry curvature is odd in momentum
space \(\Omega(k) = - \Omega( - k)\), so its integral, weighted by the
equilibrium Fermi distribution (\(f_{0}\)), must vanish because the
states at $k$ and \(- k\) are equally occupied. However, a
second-order response arises from the integral of the Berry curvature
over the non-equilibrium distribution of electrons. Since this
non-equilibrium distribution is unequal for \emph{k} and \(- k\) states,
the integral can be finite. Seminal studies on non-linear transport
effects in non-centrosymmetric systems date back to 1968
\cite{99}. More recently, E. Deyo \emph{et al.} developed
a semiclassical theory highlighting the contributions of Berry curvature
and side-jump mechanisms \cite{100}. I. Sodemann and
L. Fu \cite{101}, by generalizing a previous work on
photocurrents by J. Moore and J. Orenstein \cite{102},
demonstrated that indeed the intrinsic NLHE is tied to the
integral of the Berry curvature gradient over the occupied states,
\(D_{ab} = \int_{k}^{}{f_{0}\left( \partial_{a}\Omega_{b} \right)}\),
known as the Berry curvature dipole (BCD). Thus, the key intrinsic
property responsible for the non-dissipative NLC (i.e., NLHE) in non-magnetic
crystals with broken \(\mathcal{P}\) is the BCD. Importantly, in the DC
limit, the BCD contribution of the NLC tensor is
linearly proportional to \(\tau\), \(\sigma^{(2,0)} \propto \tau D\) (Table \ref{tab:2.3}). We
note that, for some authors, intrinsic contributions are necessarily
scattering independent, and thus the BCD is considered an extrinsic
mechanism \cite{103}. However, in the framework adopted in
this theis, extrinsic refers to mechanisms where scattering centers
have been explicitly incorporated into the calculations.

Several works \cite{101,104,105} highlighted monolayer TMDCs,
such as WTe\textsubscript{2} and MoTe\textsubscript{2}, as promising candidates for experimentally
observing the NLHE. However, \(\mathcal{P}\) and/or $C_{3v}$ symmetry
in these systems cancels the effect. Therefore, symmetry must be broken
externally, for instance, by applying out-of-plane electric fields or
through the presence of an interacting substrate. On the other hand,
even-layer WTe\textsubscript{2} naturally breaks \(\mathcal{P}\) in its pristine state.
In this system, Q. Ma \emph{et al}. \cite{37} and K. Kang \emph{et al.} \cite{38} reported the observation of the NLHE for the first time. While the results have been explained in
terms of the BCD, signatures of extrinsic contributions have also been
detected \cite{38}.

It should be noted that strong BCD is not limited to non-centrosymmetric
TMDCs. Interestingly, a strong NLC has been reported in LaAlO\textsubscript{3}/SrTiO\textsubscript{3}, originating from an orbital-mediated BCD \cite{carmine, carmine3}. Additionally, the BCD is greatly
enhanced and changes sign during topological
transitions \cite{97}, making these transitions experimentally
observable \cite{96}. This phenomenon opens the door to the
development of Berry-curvature-based memory devices.

\emph{\textbf{A.2. Extrinsic mechanisms and scaling law.} --} The study of
intrinsic and extrinsic mechanisms of the linear anomalous Hall effect
has evolved over the years due to the complexity of experimentally
disentangling the different contributions \cite{AHE}. After
numerous proposals, a refined scaling law for the anomalous Hall effect
was established \cite{106}, identifying Berry curvature as the
intrinsic mechanism, while skew-scattering and side-jump were recognized
as the extrinsic mechanisms. This understanding has been applied to
develop a scaling law for NLC soon after its experimental
observation \cite{107,108}. As discussed in the previous
section, the BCD serves as the intrinsic mechanism. By analogy with the
anomalous Hall effect, skew-scattering and side-jump mechanisms have
been proposed as extrinsic mechanisms \cite{109}. Importantly,
while the BCD induces exclusively Hall signals, the extrinsic mechanisms
also lead to Ohmic and
longitudinal NLC \cite{11,39}. Moreover, non-linear
transport is highly dependent on the crystallographic direction (see
section \ref{sec:2.1.1}). Indeed, a quantum theory has been developed that
identifies components of the NLC tensor as purely
extrinsic for certain space groups \cite{110}. Therefore, by
studying the crystallographic direction dependence and the scaling law
of the non-linear response, it may be possible to disentangle the
different contributions in experiments (Table \ref{tab:2.3}).

\begin{table}[b]
	\centering
	\caption{\textbf{Non-linear transport mechanisms.}
		Origin, contribution to NLC, NLMC and anomalous NLMC, measurement configurations (longitudinal, transverse) permitted, symmetry constraints (\(C_{3v}\), \(\mathcal{P}\), \(\mathcal{T}\), \(\mathcal{PT}\)) and \(\tau\)-scaling are displayed for each microscopic mechanism: Berry curvature dipole (BCD), skew-scattering (SK), side-jump (SJ), Berry curvature and Zeeman coupling (BC-ZC), Berry curvature polarizability (BCP), non-linear Drude (NLD) and anomalous skew-scattering (ASK). SK and SJ mechanisms contribute with multiple \(\tau\)-scalings to the non-linear transport signal (\(\tau^{0,1,2,3,4}\)). Under \(\mathcal{T}\) conditions, SK just contributes with \(\tau^{1,2,3}\) and SJ with \(\tau^{1,2}\).}
	\includegraphics[width=1\linewidth]{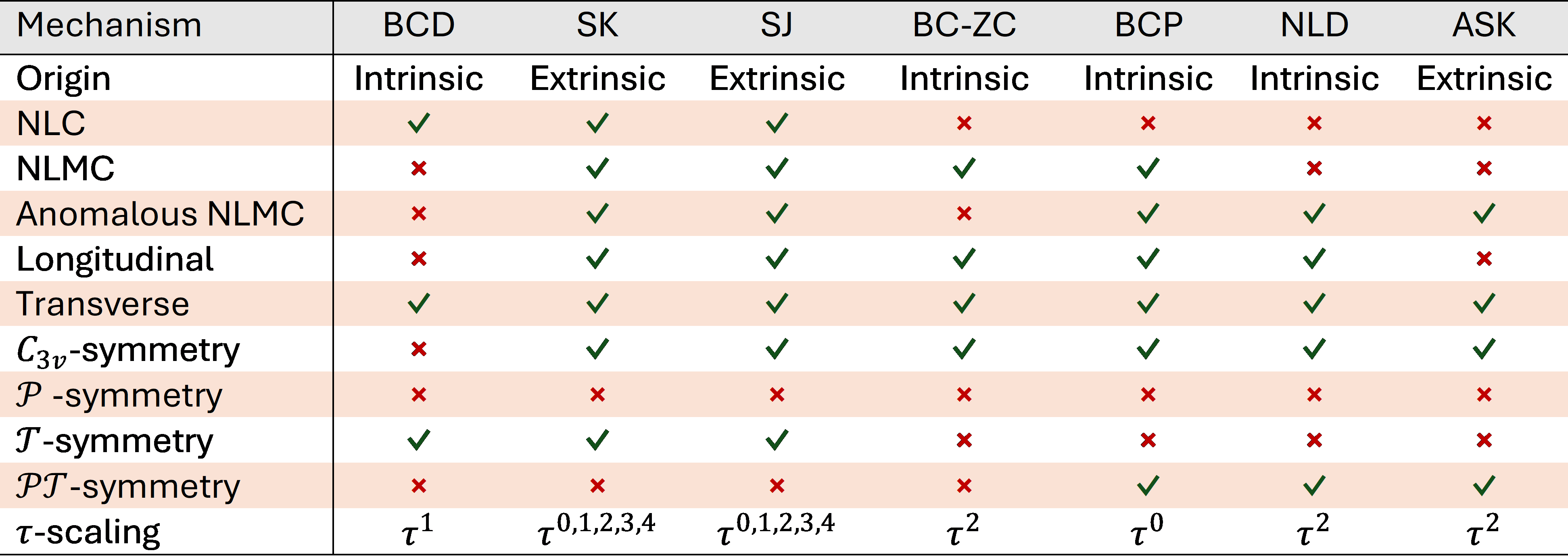}
	\label{tab:2.3}
\end{table}

\subsection{Non-linear magnetoconductivity} \label{sec:2.2.2}

A comprehensive theory of NLMC that encompasses the different physical mechanisms has still to be developed. In the case of the NLC, the intrinsic mechanism was initially proposed, followed by experimental confirmation, with later findings indicating the presence of extrinsic mechanisms. In contrast, NLMC was initially observed across different types of systems by different research communities and referred to by various names. Consequently, theories developed thus far have been mainly system- or material-specific rather than general. In particular, due to its close connection with spin textures, this effect has garnered significant interest in the spintronics community. We will explore the application of NLMC for spintronics in section \ref{sec:2.3.2}. However, as recent efforts have been made to identify overarching mechanisms, we will devote this section to summarizing them and discussing the proposed scaling law for their experimental distinction.

\emph{\textbf{B.1. Berry curvature and Zeeman coupling.} --}X. Liu \emph{et al.}
employed Berry-curvature-corrected semiclassical and Boltzmann formalism
to develop a theory of NLMC \cite{51}. They
recognize both Berry curvature and Zeeman intrinsic terms, which scale
quadratically with $\tau$. The Zeeman term can be further separated
into spin and orbital contributions, with the former requiring strong
spin-orbit coupling in the system, but not the latter. Hence, NLMC
can be utilized to explore the spin and orbital contributions within a
system. Notably, the Zeeman and Berry curvature terms apply to both longitudinal and transverse effects, which
is crucial for disentangling them from other contributions in experiments.

\emph{\textbf{B.2. Berry connection polarizability.} --} The NLMC can be
understood as a NLC in a system perturbed by \(B\). In analogy to the
intrinsic linear anomalous Hall effect, the anomalous velocity of
electrons is given by \(v^{A} = E \times \ \Omega\), but to study the
non-linear response at the second-order \(E^{2}\), the electric field
correction of the Berry curvature (\(\Omega^{E}\)) should be considered,
\(v^{A} = E \times \Omega^{E}\). This correction has been calculated by
Y. Gao \emph{et al.} showing that
\(\Omega^{E} = \nabla_{k} \times \mathcal{A}^{E}\), where
\(\mathcal{A}_{a}^{E} = \widetilde{G_{ab}}E_{b}\) is the field-corrected
Berry connection and \(\widetilde{G_{ab}}\) is the positional shift or
Berry connection polarizability (BCP) \cite{92}. Note that, here, the BCP is
defined with respect to the \emph{B}-perturbed band structure. By
expressing all quantities in the original unperturbed eigenstates,
Y.-H. Huang \emph{et al.} demonstrated that the
non-linear response is given by the integral over the occupied states of
two products: the band velocity times the spin susceptibility of BCP and
the \emph{k}-space dipole of BCP times the spin magnetic moment \cite{111}. Later
results also suggested an orbital contribution \cite{112}.
Importantly, the \emph{B}-perturbed BCP contribution to the NLMC (\(\sigma_{ijkl}^{(2,1)}\)) is reported to be
\(\tau\)-independent and purely transverse \cite{111,112}.
However, regarding the latter, there are some discrepancies within the
community (for more details, see section \ref{sec:2.2.3}.1).

\emph{\textbf{B.3. Extrinsic mechanisms and scaling law.} --} Experiments have
highlighted the significance of extrinsic contributions to the NLMC. In
this regard, a scaling law has been developed for \(\mathcal{T}\)-odd
non-linear transport in magnetic materials \cite{113}. In
these systems, the internal magnetic vectors play a role similar to that
of \(B\) in the NLMC. However, \(B\) induces additional orbital magnetic
field contributions. To address this difference and obtain a proper
scaling law for NLMC, an experimental study has modified the scaling law
for \(\mathcal{T}\)-odd non-linear transport in magnetic systems
including the additional orbital magnetic field contributions. By
applying the proposed scaling law, the dominant microscopic mechanisms
were identified for the case of chiral Te \cite{Manu4}.

\subsection{Anomalous non-linear magnetoconductivity} \label{sec:2.2.3}

As discussed in section \ref{sec:2.1.2}.5, non-centrosymmetric magnetic materials,
besides breaking \(\mathcal{P}\), also break \(\mathcal{T}\) giving rise
to \(\mathcal{T}\)-odd mechanisms governed by the magnetization (\(M\))
or Néel (\(N\)) vector. Interestingly, there are systems that break both
\(\mathcal{P}\) and \(\mathcal{T}\), but not \(\mathcal{PT\ }\)symmetry.
The preservation of \(\mathcal{PT}\) symmetry suppresses most of the
mechanisms discussed in previous sections, such as the BCD,
skew-scattering, and side-jump (see Table \ref{tab:2.3}). Therefore, systems that
preserve \(\mathcal{PT\ }\)symmetry while breaking both \(\mathcal{P}\)
and \(\mathcal{T}\) are of special interest for studying the mechanisms
covered in this section.

\emph{\textbf{C.1. Berry connection polarizability and non-linear Drude.} --}
While we have already discussed the role of the BCP in the NLMC, we will
now review that, in magnetic systems, the BCP contributes to non-linear
transport even in the absence of \(B\). As detailed in section \ref{sec:2.2.2}.2,
the Berry connection receives a correction from the applied electric
field \(\Omega^{E} = \nabla_{k} \times \mathcal{A}^{E}\), where
\(\mathcal{A}_{a}^{E} = G_{ab}E_{b}\)\hspace{0pt} \cite{92}.
Here, since we are not considering any \(B\), the BCP (\(G_{ab}\)) is
defined relative to the unperturbed band structure. Following the same
strategy used for the NLMC, H. Liu \emph{et al.} \cite{103} and
C. Wang \emph{et al.} \cite{114} calculated the
\(\tau\)-independent contribution to the non-linear transverse transport
in magnetic systems, showing that it stems from the integral over the
occupied states of the BCP dipole moment. Subsequent studies based on
the quantum response theory have suggested that this
\(\tau\)-independent term may also play a role in longitudinal
transport \cite{115,116}. This hypothesis was substantiated by
a recent investigation that explicitly demonstrated the equivalence
between semiclassical and quantum response theories up to second order
in the electric field \cite{117}.

We want to highlight that this \(\tau\)-independent mechanism has been
referred to in the literature as the band-normalized quantum
metric \cite{12,90,115,116, giacomo, carmine4}. The quantum metric and Berry
curvature represent the real and imaginary parts of the quantum
geometric tensor, respectively. In its definition, the BCP contains an
extra energy denominator compared with the quantum metric, which
generically makes it a different, non-geometric quantity. While there
might be an approximate connection between the two when only two bands
are involved, the BCP correction remains a more faithful name to the
meaning of this quantity, and we have thus used it in this way in this
review. This is of course not withstanding the legitimate high interest
of the quantum metric in condensed matter systems and quantum
transport, which may be found to be connected in other ways in the
future.

Additionally, a non-linear Drude term has also been
reported \cite{115,118}. This term originates from the
skewness of the energy dispersion, which is generally non-zero when both
\(\mathcal{P}\) and \(\mathcal{T}\) are separately broken. It only
depends on the band dispersion of the states at the Fermi surface, and
on \(\tau\). While the linear Drude term is
proportional to \(\tau\), the non-linear Drude is proportional to
\(\tau^{2}\), which makes it \(\mathcal{T}\)-odd but
\(\mathcal{PT}\)-even. The non-linear Drude term in magnetic systems is
equivalent to the Zeeman coupling term in the NLMC (see section \ref{sec:2.2.2}.1).

\emph{\textbf{C.2. Extrinsic mechanism and scaling law.} --} \(\mathcal{PT}\)
symmetry cancels out the BCD and the conventional
skew-scattering mechanisms discussed in the previous
sections \cite{119}. However, in magnetic systems, other
extrinsic mechanisms may still emerge even under \(\mathcal{PT}\)
symmetry conditions. For example, an anomalous skew-scattering mechanism
has been proposed, where Berry curvature and skew-scattering work in
tandem \cite{119}. A scaling law for \(\mathcal{T}\)-odd
non-linear transport in magnetic systems (anomalous NLMC) has been
proposed \cite{113}. However, non-centrosymmetric magnetic
systems generally also exhibit a broad range of mechanisms contributing
to \(\mathcal{T}\)-even transport (NLC), resulting in numerous
possibilities. Therefore, while several scaling laws have been
proposed \cite{113,118,120}, it is crucial to also consider
the different symmetry constraints of these mechanisms to
disentangle them in experiments (see Table \ref{tab:2.3}).

\section{Applications} \label{sec:2.3}

Besides their importance in accessing intriguing physical properties of studied systems, non-linear transport effects have several significant applications. On one hand, the same symmetry requirements that give rise to non-linear transport phenomena also result in electronic-band splitting and the emergence of spin and orbital textures. As a result, non-linear effects can be leveraged to probe the angular momentum texture of a system by purely electrical means. On the other hand, non-centrosymmetric systems exhibit diode-like behavior within a single material, overcoming some of the key limitations of conventional rectifiers, and making them an attractive alternative for energy harvesting applications. In this section, these two main applications will be discussed.

\subsection{Spintronics and orbitronics} \label{sec:2.3.1}

Spintronics and orbitronics are branches of electronics that harness,
respectively, the spin and orbital angular momentum of electrons, in
addition to their charge. They promise significantly lower power
consumption \cite{121,122} although, to expand their
applications, it is crucial to develop systems that enable accurate
generation, control, and detection of electron angular momentum. In this
regard, non-centrosymmetric systems may offer an outstanding platform.

\begin{figure}[t]
	\centering
	\includegraphics[width=1\linewidth]{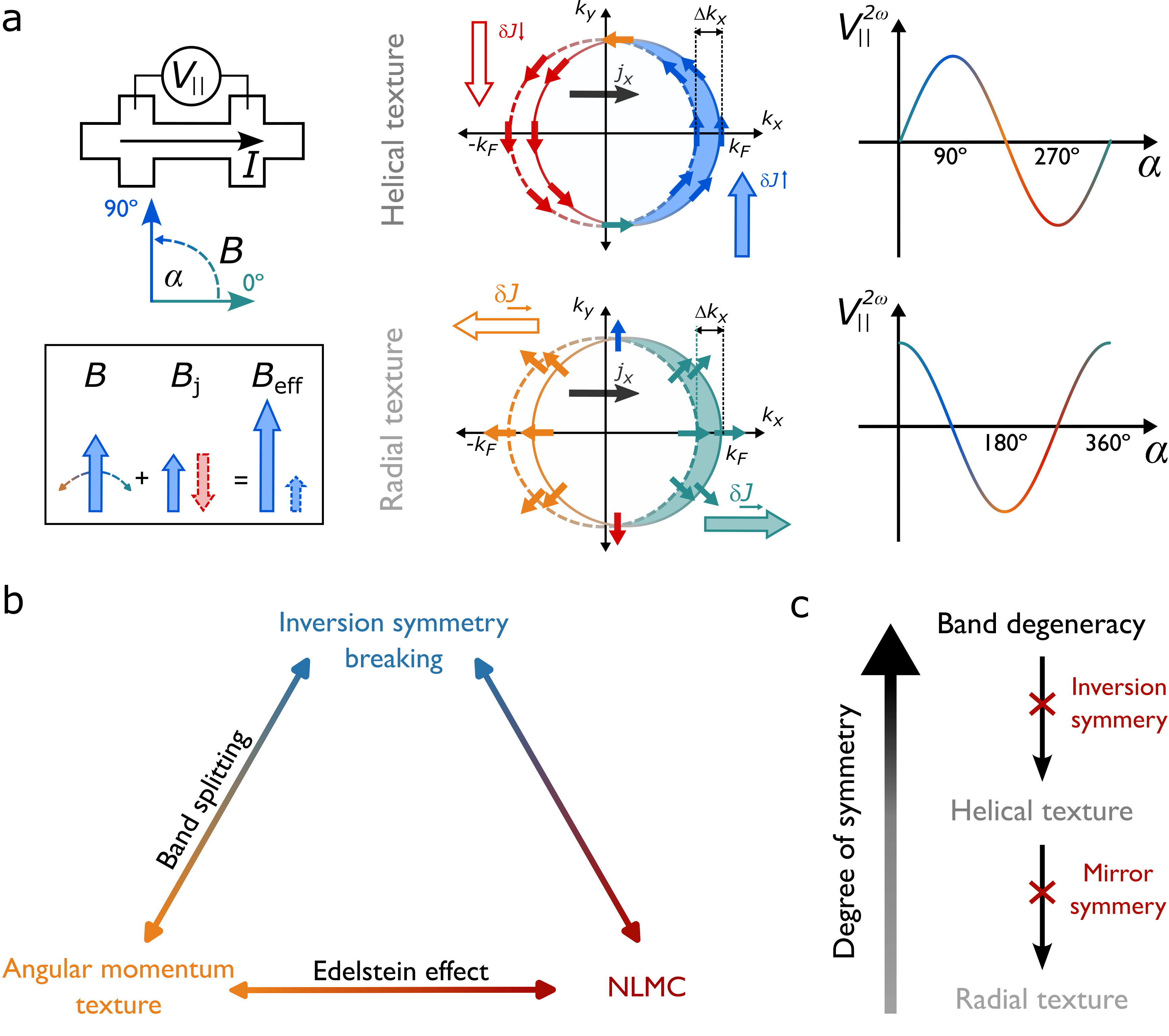}
	\caption{\textbf{Connection between NLMC and angular
			momentum textures.}
		\textbf{a}, Measurement configuration to probe
		helical and radial textures. There is a representation of both textures,
		which included the current-induced angular momentum accumulation
		(\(\delta J\)) in response to a current (\(j_{x}\)) through the
		Edelstein effect. In the black square, the effective field (\(B_{eff}\))
		as a result of the external (\(B\)) and current-induced field
		(\(B_{j}\)) is displayed. Finally, the expected \(\alpha\)-angle
		dependence of the second-harmonic parallel voltage
		(\(V_{\parallel}^{2\omega}\)) is included for each texture. \textbf{b},
		The relationship between inversion symmetry (\(\mathcal{P}\)) breaking, angular momentum
		texture and NLMC is sketched. \textbf{c}, The different symmetries that
		must be broken in order to allow helical and radial textures are
		displayed.}
	\label{fig:2.3}
\end{figure}

In systems with broken \(\mathcal{P}\), spin-orbit coupling (SOC) has
traditionally been considered necessary for band splitting, leading to
the formation of spin textures \cite{123,124},
\(E_{S}(k) \neq E_{- S}(k)\). However, recent studies suggest that, in
the absence of SOC, orbital angular momentum can similarly lift band
degeneracy, generating orbital textures
instead \cite{125,126}, \(E_{L}(k) \neq E_{- L}(k)\). When an
electric field is applied, V. Edelstein predicted that these
textures force transport electrons to align their angular momentum in a
specific direction, resulting in angular momentum
accumulation \cite{127}. This accumulation can be intuitively
understood as a current-induced internal magnetic field. Consequently,
an external magnetic field parallel or antiparallel to the
current-induced internal magnetic field will lead to different effective
fields, causing a resistance difference for opposite current directions
(Fig. \ref{fig:2.3}.a). This phenomenon is none other than the NLMC and provides a
simple method to probe angular momentum textures through purely
electrical means \cite{26,29}.

Besides this intuitive interpretation, two detailed theories have been
developed to explain the longitudinal NLMC in terms of the spin texture
of TIs \cite{26,30} and 2DEGs \cite{29}. The first
approach is based on the conversion of a non-equilibrium spin current
into a charge current under the application of an external magnetic
field \cite{26}, while the second focuses on the interplay
between current-induced spin polarization and scattering processes due
to inhomogeneities of spin-momentum locking \cite{29,30}. The
former has been generalized to also explain the transverse
NLMC \cite{34}. While these microscopic descriptions offer an
alternative approach to explaining NLMC, the mechanisms discussed in
section \ref{sec:2.2.2} provide a more direct approach (Fig. \ref{fig:2.3}.b). Nonetheless,
angular momentum textures exist and have been directly observed through
spin-resolved angle-resolved photoemission
spectroscopy \cite{sakano,128}. Therefore, both strategies are
valid, and by combining them, it is possible to exploit symmetry
analysis to predict angular momentum textures and use NLMC measurements
to probe them.

In particular, when \(\mathcal{P}\) is broken while mirror symmetry is
preserved, spin textures perpendicular to the mirror planes emerge (Fig. \ref{fig:2.3}.c). This is the classic example of 2DEGs \cite{27,29,31} and
TIs \cite{26,67,68}. In these systems, when surface symmetry
is maintained, two perpendicular mirror planes constrain the spins to
the surface, resulting in helical spin
textures \cite{29,31,67}. However, if surface symmetry is
disrupted and only one mirror plane is retained, out-of-plane spin
components emerge \cite{26,27}. Importantly, as shown in Table
\ref{tab:2.1}, polar groups with two mirror planes (\emph{mm2}, \emph{4mm}, and
\emph{6mm}) enable NLMC in only one configuration with the magnetic
field aligned perpendicular to the current and both mirror planes,
consistent with a helical spin texture. Conversely, polar groups with a
single mirror plane (\emph{m} and \emph{3m}) allow NLMC in two
configurations with the magnetic field oriented in both directions
perpendicular to the current, also in agreement with the spin texture of
those systems. Furthermore, when all mirror symmetries are broken,
radial spin textures arise (Fig. \ref{fig:2.3}.c), as seen in chiral systems like
trigonal Te \cite{sakano}. Indeed, chiral groups permit NLMC with
magnetic fields parallel to the current (Table \ref{tab:2.1}), aligning with the
radial spin texture. Therefore, the same symmetry
conditions that give rise to angular momentum textures also allow NLMC
effects. This interplay can be harnessed to probe angular momentum
textures through non-linear transport measurements. Moreover, following
the symmetry analysis, it is possible to engineer systems with specific
angular momentum textures \cite{130}, providing versatile
platforms for applications in spintronics and orbitronics.

In this section, another physical effect should be mentioned, possibly
related to NLMC, which is the so-called chirality induced spin
selectivity (CISS). It has attracted much attention as a way to
generate large spin polarization in chiral molecular systems. However,
the phenomenology is largely constrained to materials with very low
electrical conductivity, and the main mechanisms behind the observed
experiments remain unclear \cite{131}.

\subsection{Energy harvesting} \label{sec:2.3.2}

In recent years, there has been an exponential increase in ambient
RF radiation, accompanied by the rapid proliferation of
IoT devices \cite{joule}. Consequently, wireless rectifying antennas
(rectennas), which can effectively harvest ambient RF radiation to power
IoT devices, have emerged as a critical technology for reducing global
energy consumption \cite{RFharvesting}. State-of-the-art rectennas
combine an antenna to collect the RF radiation and convert it into an AC
current with a rectifier that converts the AC signal into a DC current.
Schottky diodes are the most widely used alternative for the
rectification module. However, they exhibit a threshold input voltage,
referred to as thermal voltage limitation, below which the device fails
to operate effectively \cite{RF,135}. This makes the inclusion
of an antenna module to enhance the incoming signal essential, leading
to a substantial increase in the overall size of the rectenna and
consequently constraining its applicability in microscale devices. In
this context, non-centrosymmetric systems may offer an exceptional
solution.

\begin{figure}[t]
	\centering
	\includegraphics[width=1\linewidth]{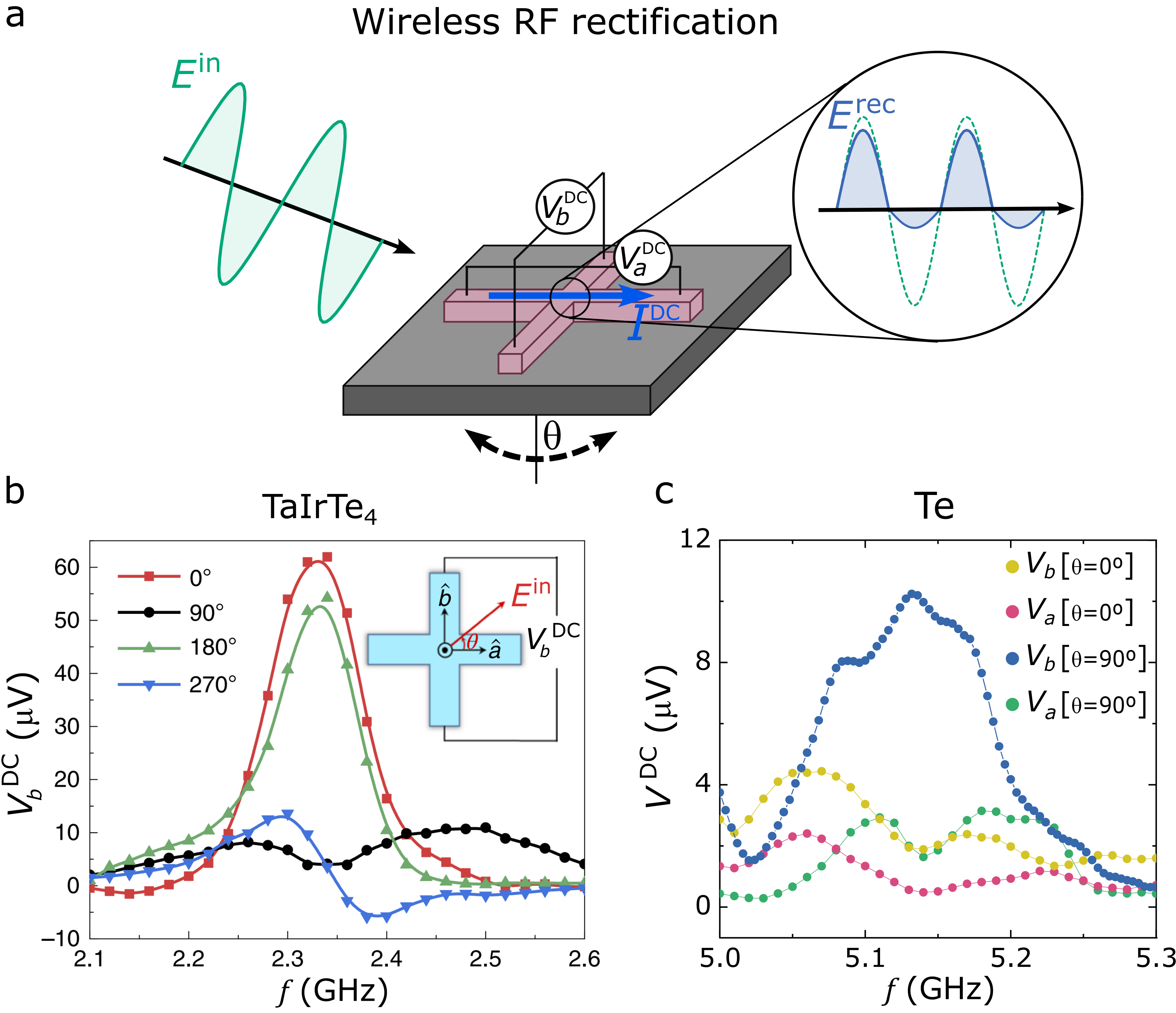}
	\caption{\textbf{Wireless RF rectification.}
		\textbf{a}, An incident RF wave (\(E^{in}\)) generates an asymmetric
		electric field (\(E^{rec}\)) due to the non-linear transport
		characteristics of non-centrosymmetric systems, resulting in a DC
		current (\(I^{DC}\)) within the material. \textbf{b}, \textbf{c},
		Rectified DC voltage (\(V^{DC}\)) as a function of frequency (\(f\)) in \textbf{b}, TaIrTe\textsubscript{4} and \textbf{c}, Te. The signal is
		maximum when the incident polarization is \textbf{b}, transverse (\textbf{c}, parallel) to the voltage detection direction in agreement
		with the non-linear transport characteristics of \textbf{b}, TaIrTe\textsubscript{4} (\textbf{c}, Te). Figure adapted with permission from \textbf{b}, Ref. \cite{72} (Springer Nature Limited); and \textbf{c}, Ref. \cite{Manu3} (Wiley).}
	\label{fig:2.4}
\end{figure}

As discussed throughout the chapter, non-centrosymmetric systems
exhibit intrinsic non-linear transport characteristics. Therefore, when
an incoming RF wave (\(E^{in}\)) is absorbed by the material, an
asymmetric electric field (\(E^{rec}\)) is generated. The temporal
integration of \(E^{rec}\) results in a net DC current
(\(I^{DC}\)) within the material \cite{RF,135} (Fig. \ref{fig:2.4}.a).
Importantly, the physics of the rectification mechanism is completely
different from that of Schottky diodes
and is therefore not restricted by the thermal voltage limitation. In
low-power applications, this allows for the elimination of the antenna
module, dramatically reducing the dimensions of the rectenna to the
microscale or even nanoscale.

The practical implementation of RF energy harvesting using a
non-centrosymmetric system was first demonstrated in TaIrTe\textsubscript{4}, where
symmetry is disrupted at the surface level \cite{72}. Since
the rectification is based on the transverse NLC, \(I^{DC}\) is recorded
transverse to the polarization direction of \(E^{in}\) (Fig. \ref{fig:2.4}.b).
Subsequent experiments in magnetic \cite{12} and
polar \cite{56} systems also demonstrated wireless
rectification in a transverse configuration. Additionally, chiral Te
exhibited RF rectification in a longitudinal configuration, with
\(V^{DC}\) measured parallel to the polarization direction of
\(E^{in}\), consistent with its longitudinal NLC \cite{Manu3}
(Fig. \ref{fig:2.4}.c). Recently, Bi\textsubscript{2}Te\textsubscript{3} has exhibited rectification in both
longitudinal and transverse configurations at frequencies as high as
27.4 GHz, which is crucial for the next generation of 5G frequency
bands \cite{137}. In Table \ref{tab:2.4}, the main reports in literature
are summarized. Notably, in all these experiments, an antenna module to
collect and enhance the incoming signal was not required, achieving full
energy harvesting performance at microscale dimensions. Although the
efficiencies remain orders of magnitude below those of state-of-the-art
devices combining an antenna with a Schottky diode \cite{RFcomp1, RFcomp2},
these results pave the way for the development of self-powered
microscale devices (see section \ref{sec:2.4}). 

\begin{table}[t]
	\centering
	\caption{\textbf{Summary of the main experimental
			results on wireless RF rectification.}
		The table includes the
		non-centrosymmetric material where the effect is reported, mechanism,
		temperature (\emph{T}), frequency (\(f\)), incident power (\(P^{in}\)),
		output power (\(P^{out}\)), and efficiency (\(\eta = P^{out}/P^{in}\)).}
	\includegraphics[width=1\linewidth]{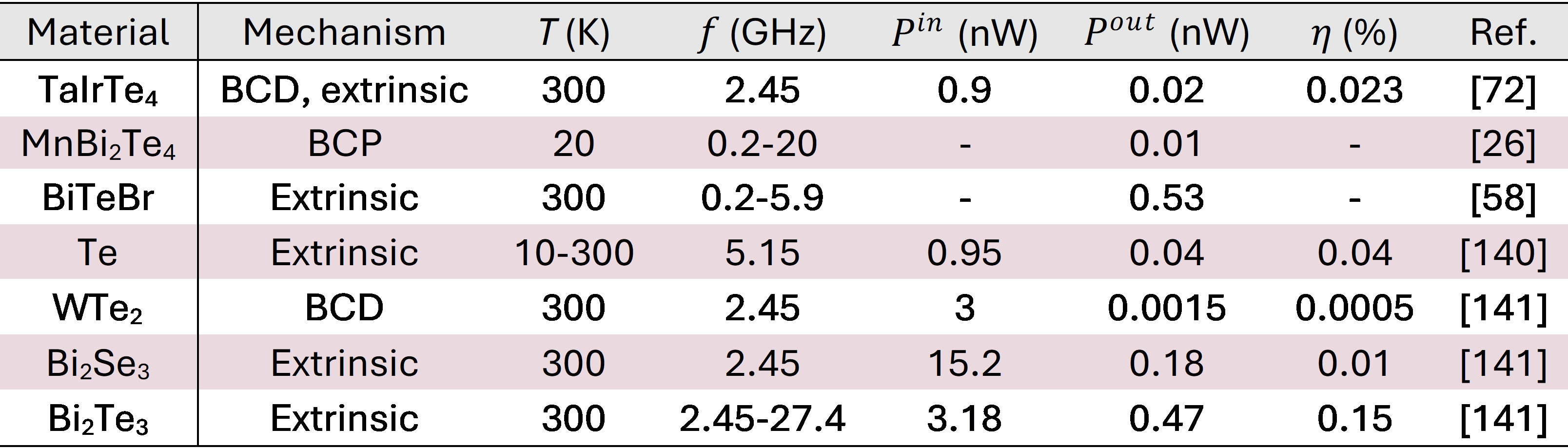}
	\label{tab:2.4}
\end{table}

At higher frequencies, in the visible regime, an equivalent effect known
as the bulk photovoltaic effect (BPVE) has also been observed.
State-of-the-art photovoltaic cells are based on \emph{p-n} junctions,
whose efficiencies are limited by the Shockley-Queisser limit. Since the
BPVE is not subject to this limit, it has attracted considerable
attention for self-powered harvesting applications \cite{138}.
Although the underlying principle of the effect is the same as wireless
RF rectification and subject to the same symmetry constraints (see Table
\ref{tab:2.1}), the higher frequencies and energies of visible light result in
electronic band transitions, different transport regimes, additional
microscopic mechanisms, and thermal effects. Therefore, it falls outside
the scope of this review \cite{139,140}.

\section{Conclusion} \label{sec:2.4}

In conclusion, the field of non-linear transport effects in
non-centrosymmetric materials is rapidly advancing, driven by both
fundamental insights and broad application potential. Despite
significant progress, key challenges and opportunities remain to be
addressed.

i) For application purposes, there is significant interest in achieving
stronger non-linear responses. Intrinsic mechanisms, such as the Berry
curvature dipole \cite{101} and Berry connection
polarizability \cite{111}, are well understood, enabling the
identification of key material characteristics that enhance non-linear
transport. Although non-centrosymmetric materials for electronic
transport are limited, material engineering has shown promise in
overcoming this limitation. Among the various approaches, van der Waals
heterostructures, offering the flexibility to stack \cite{39}
and twist \cite{40} different layers, stand out as an ideal
platform \cite{141}.

At the same time, experimental studies have revealed the significant and
often dominant role of extrinsic mechanisms \cite{39, Manu2, Manu4}, which
remain less understood. Advancing theoretical and experimental research
in this area could be crucial for achieving substantial enhancements in
non-linear output.

ii) While initial experiments have primarily focused on output
enhancement, non-centrosymmetric systems offer a broad range of
possibilities. One particularly intriguing avenue is the exploration of
non-volatile responses for memory applications. Non-linear transport
effects enable direct reading of magnetization, Néel, and polarization
vector orientations \cite{103}, opening new pathways for
device functionality. Preliminary studies on
ferromagnetic \cite{95},
antiferromagnetic \cite{12,90,93,95}, and
ferroelectric \cite{73} systems have already demonstrated
reconfigurable and permanent non-linear responses, yet significant
potential remains to be explored in this field.

iii) Wireless rectification has been demonstrated at room
temperature \cite{72}, but its efficiency still falls behind
that of state-of-the-art Schottky diode devices. The relationship
between non-linear conductivity and non-linear resistivity shows that
higher rectifier resistance leads to stronger non-linear output voltages
(section \ref{sec:2.1.1}). However, it is important to note that wireless
rectification devices must transfer the harvested
power \cite{5}. This transfer process limits the rectifier
resistance, as it becomes inefficient if the receiving device is more
conductive \cite{Manu3}. To overcome this limitation, the
mechanism behind rectification in non-centrosymmetric systems may also
provide the solution, as it is not constrained by the band gap. This
allows for engineering an optimal balance between resistivity and high
absorption, enabling both strong rectification output and efficient
power transfer. Moreover, the working frequency spectrum is not a
fundamental limitation, meaning the devices can be optimized to operate
across a broad spectrum, from radio to far-infrared, including
visible and THz frequencies.

iv) In the fields of spintronics and orbitronics, achieving magnetic
states writing and reading with high precision and low power consumption
is critical. In this regard, non-centrosymmetric materials present a
promising alternative, as their symmetry can be engineered to drive and
detect the targeted angular momentum orientations. This review provides
a framework for identifying such symmetries, which may guide the
development of the next generation of spin-orbit torque and
magneto-electric spin-orbit devices. Indeed, early applications of
non-centrosymmetric systems in these devices have demonstrated promising
performance \cite{130,142}.

\appendix
\section{From conductivities to resistivities}

\label{s:note}
As we have discussed through the review, non-linear transport effects are theoretically described in terms of conductivities. However, it is common to experimentally measure resistivities. In this appendix, we will obtain the relationship between them up to second-order in electric field ($E$) and magnetic field ($B$). In general, the full current response is given by:
\begin{multline}
	j_{i} = \sigma_{ij}^{(1,0)}E_{j}^{\omega} + \sigma_{ijk}^{(1,1)}E_{j}^{\omega}B_{k} + \sigma_{ijkl}^{(1,2)}E_{j}^{\omega}B_{k}B_{l} + \\ + \sigma_{ijk}^{(2,0)}E_{j}^{\omega}E_{k}^{\omega}  + \sigma_{ijkl}^{(2,1)}E^{\omega}_{j}E_{k}^{\omega}B_{l},
	\label{eq:2.2}
\end{multline}
where we have considered the Ohmic conductivity ($ \sigma_{ij}^{(1,0)} $), the Hall conductivity ($ \sigma_{ijk}^{(1,1)} $), the magnetoconductivity ($\sigma_{ijkl}^{(1,2)}$), the non-linear conductivity ($  \sigma_{ijk}^{(2,0)} $), and the non-linear magnetoconductivity ($ \sigma_{ijkl}^{(2,1)} $). We define the analog coefficients in resistivitiy as:
\begin{multline}
	E_{i} = \rho_{ij}^{(1,0)}j_{j}^{\omega} + \rho_{ijk}^{(1,1)}j_{j}^{\omega}B_{k} +  \rho_{ijkl}^{(1,2)}j_{j}^{\omega}B_{k}B_{l} + \\ 
	+ \rho_{ijk}^{(2,0)}j_{j}^{\omega}j_{k}^{\omega}  + \rho_{ijkl}^{(2,1)}j^{\omega}_{j}j_{k}^{\omega}B_{l} .
	\label{eq:2.3}
\end{multline}
Plugging Eq. (\ref{eq:2.3}) back into Eq. (\ref{eq:2.2}) and equating order by order, we can relate conductivities to resistivities. 

At zero-order in $ B $ and first-order in $ E $, we find:
\begin{equation}
	\sigma^{(1,0)}_{ij}\rho^{(1,0)}_{jk} = \delta_{ik}.
\end{equation}
To first-order in $ B $ and $ E $, we obtain:
\begin{equation}
	\sigma_{ij}^{(1,0)}\rho^{(1,1)}_{jkl}j_{k}^{\omega}B_{l} + \sigma_{ijk}^{(1,1)}\rho^{(1,0)}_{jm}j_{m}^{\omega}B_{k} = 0,
\end{equation}
which results:
\begin{equation}
	\rho^{(1,1)}_{ijk} = -\rho_{il}^{(1,0)}\sigma_{lmk}^{(1,1)}\rho^{(1,0)}_{mj}.
\end{equation}
To second-order in $ B $ and first-order in $ E $, we get:
\begin{equation}
	\sigma_{ij}^{(1,0)}\rho^{(1,2)}_{jklm}j_{k}^{\omega}B_{l}B_{m} + \sigma_{ijkl}^{(1,2)}\rho^{(1,0)}_{jn}j_{n}^{\omega}B_{k}B_{l} = 0,
\end{equation}
which gives:
\begin{equation}
	\rho^{(1,2)}_{ijkl} = -\rho^{(1,0)}_{il}\sigma_{lmkl}^{(1,2)}\rho^{(1,0)}_{mj}.
\end{equation}
To second-order in $ E $ and zero-order in $B$, we find:
\begin{equation}
	\sigma^{(1,0)}_{ij}\rho_{jkl}^{(2,0)}j_{k}^{\omega}j_{l}^{\omega} + \sigma^{(2,0)}_{ijk}\rho^{(1,0)}_{jm}j_{m}^{\omega}\rho^{(1,0)}_{kn}j_{n}^{\omega} = 0,
\end{equation}
which is equivalent to:
\begin{equation}
	\rho_{pkl}^{(2,0)}j_{k}^{\omega}j_{l}^{\omega} = -\rho^{(1,0)}_{pi}\sigma^{(2,0)}_{ijk}\rho^{(1,0)}_{jm}\rho^{(1,0)}_{kn}j_{m}^{\omega}j_{n}^{\omega}.
\end{equation}
Now, relabeling indices:
\begin{equation}
	\rho_{ijk}^{(2,0)} = -\rho^{(1,0)}_{il}\sigma^{(2,0)}_{lmn}\rho^{(1,0)}_{mj}\rho^{(1,0)}_{nk}.
\end{equation}
Finally, to second-order in $ E $ and first-order in $ B $, we have:
\begin{multline}
	\sigma_{ij}^{(1,0)}\rho^{(2,1)}_{jklm}j_{k}^{\omega}j_{l}^{\omega}B_{m} + \sigma^{(2,0)}_{ijk}(\rho^{(1,0)}_{jl}j_{l}^{\omega}\rho^{(1,1)}_{kmn}j_{m}^{\omega}B_{n} + \\ 
+ \rho^{(1,1)}_{jmn}j_{m}^{\omega}B_{n}\rho^{(1,0)}_{kl}j_{l}^{\omega}) +  \sigma^{(1,1)}_{ijk}\rho_{jnm}^{(2,0)}j_{n}^{\omega}j_{m}^{\omega}B_{k} + \\ + \sigma_{ijkl}^{(2,1)}\rho^{(1,0)}_{jm}j_{m}^{\omega}\rho_{kn}j_{n}^{\omega}B_{l} = 0,
\end{multline}
which gives after some relabeling:
\begin{multline}
	\rho^{(2,1)}_{abcd} = \left[\left(\sigma^{(2,0)}_{ijk}\sigma_{lmd}^{(1,1)} + \sigma^{(2,0)}_{imk}\sigma_{ljd}^{(1,1)} + \sigma^{(2,0)}_{kjm}\sigma_{ild}^{1,1}\right)\rho^{(1,0)}_{kl}\right.  \\ 
	-\left. \sigma^{(2,1)}_{ijmd}\right]\rho^{(1,0)}_{ia}\rho^{(1,0)}_{jb}\rho^{(1,0)}_{mc}.
\end{multline}
In addition to the non-linear magnetoconductivity contribution ($\sigma^{(2,1)}_{ijkl}$), the non-linear magnetoresistivity ($\rho^{(2,1)}_{ijkl}$) also includes a contribution arising from the interplay between non-linear conductivity ($\sigma^{(2,0)}_{ijk}$) and the ordinary Hall effect ($\sigma^{(1,1)}_{ijk}$). However, it emerges at a higher order in Ohmic resistivity ($\rho^{(1,0)}_{ij}$). In summary, the relationship between resistivities and conductivities is given by:
\begin{flalign}
	&\rho^{(1,0)}_{ij} = (\sigma^{(1,0)}_{ij})^{-1}, &&\\
	&\rho^{(1,1)}_{ijk} = -\rho^{(1,0)}_{il}\sigma_{lmk}^{(1,1)}\rho^{(1,0)}_{mj}, &&\\
	&\rho^{(1,2)}_{ijkl} = -\rho^{(1,0)}_{il}\sigma_{lmkl}^{(1,2)}\rho^{(1,0)}_{mj}, &&\\
	&\rho^{(2,1)}_{abcd} = \left[\left(\sigma^{(2,0)}_{ijk}\sigma_{lmd}^{(1,1)} + \sigma^{(2,0)}_{imk}\sigma_{ljd}^{(1,1)} + \sigma^{(2,0)}_{kjm}\sigma_{ild}^{1,1}\right)\rho^{(1,0)}_{kl}\right. && \nonumber \\ 
	& -\left. \sigma^{(2,1)}_{ijmd}\right]\rho^{(1,0)}_{ia}\rho^{(1,0)}_{jb}\rho^{(1,0)}_{mc}.
\end{flalign}

	\begin{acknowledgments}
	This work is supported by the Spanish MICIU/AEI/10.13039/501100011033
	and by ERDF/EU (Projects No.\ PID2021-122511OB-I00, No.\
	PID2021-128760NB-I00, No.\ PID2021-129035NB-I00, and ``Maria de Maeztu''
	Units of Excellent Programme No.\ CEX2020-001038-M). It is also supported
	by MICIU and by the European Union NextGenerationEU/PRTR-C17.I1, as well
	as by IKUR Strategy under the collaboration agreement between Donostia
	International Physics Center and CIC nanoGUNE on behalf of the
	Department of Education of the Basque Government. M.S.-R. acknowledges
	support from La Caixa Foundation (No.\ 100010434) with code
	LCF/BQ/DR21/11880030. M.G. thanks support from the ``Ramón y Cajal''
	Programme by the Spanish MICIU/AEI and European Union
	NextGenerationEU/PRTR (Grant No.\ RYC2021-034836-I).
	\end{acknowledgments}

\bibliographystyle{apsrev4-2}
\bibliography{Bibliography}
	
\end{document}